\documentclass[amsmath,amssymb,aps,pre,reprint,groupedaddress]{revtex4-1}
\usepackage{graphicx}
\usepackage{float}
\usepackage[caption=false]{subfig}
\usepackage{tensor}
\newcommand{\mathsym}[1]{{}}
\newcommand{\unicode}[1]{{}}
\usepackage[usenames]{color}

\begin{document}

\title{Elasticity in Curved Topographies: Exact Theories and Linear Approximations}


\author{Siyu Li}
\email[]{siyu.li@email.ucr.edu}
\author{Roya Zandi}
\affiliation{Department of Physics and Astronomy,
   University of California, Riverside, California 92521, USA}
\author{Alex Travesset}

\affiliation{Department of Physics and Astronomy, Iowa State University and Ames lab, Ames, IA, 50011}


\date{\today}

\begin{abstract}

Almost all available results in elasticity on curved topographies are obtained within either a small curvature expansion or an empirical covariant generalization that accounts for screening between Gaussian curvature and disclinations. In this paper, we present a formulation of elasticity theory in curved geometries that unifies its underlying geometric and topological content with the theory of defects. The two different linear approximations widely used in the literature are shown to arise as systematic expansions in {\em reference} and {\em actual} space. Taking the concrete example of a 2D crystal, with and without a central disclination, constrained on a spherical cap, we compare the exact results with different approximations and evaluate their range of validity. We conclude with some general discussion about the universality of non-linear elasticity.

\end{abstract}

\pacs{81.16.Fg, 81.16.Dn, 81.07.Bc, 81.05.Kf}

\maketitle

\section{Introduction}

There are many examples of 2D crystals on curved spaces, including colloids absorbed on a spherical surface~\cite{BauschMe2003,Sanaz2018}, negative curvature~\cite{IrvineVitelliChaikin2010} at oil-water interface, virus shells~\cite{Lidmar2003,Wagner2015,ning2016vitro} and colloids mixtures~\cite{VernizziOlveradelaCruz2007}, just to name a few. The uniqueness of these problems arises from the subtle but profound relation between geometry and topology. 

The equilibrium structure of two-dimensional ordered structures on the surfaces of non-zero Gaussian curvature is dictated by the presence and arrangement of defects such as dislocations and disclinations.  The energetically forbidden defects in flat surfaces become ubiquitous on curved substrates; nevertheless, their presence gives rise to equilibrium structures that include finite stresses. The standard theory of elasticity~\cite{LandauElasticityBook} is unwieldy to investigate the interplay of the defects and geometry and, often, is not the most suitable starting point for these problems. In fact, in order to satisfy topological constraints, somewhat uncontrolled approximations need to be considered. 

In this paper we develop a geometric theory for elasticity that incorporates topological constraints exactly, thus allowing to calculate the stress and strain in a curved surface and analyze different approximations employed in the literature.  Examples that will be discussed include five-fold disclinations in a triangular lattice in the regions of constant positive Gaussian curvature, see Fig.\ref{Fig:Ref_vsTarg}. 

The organization of the paper is as follows: First, in Sect.~\ref{Sect_Conceptual} we present different approximations employed in literature to solve elasticity equations and provide a conceptual discussion of our approach, which is developed in Sect.~\ref{Sect_Formalism}. As an example, the case of a spherical cap, with or without a central disclination and the derivation of all their relevant analytical formulas are presented in Sect.~\ref{Sect_Results}. Explicit comparisons between the different approximations and the exact results are presented in Sect.~\ref{Sect_Discussion}. Some general conclusions are presented in Sect.~\ref{Sect_Conclusions}. More technical/mathematical developments are deferred to the appendices, where we have made a special effort in providing all the detail necessary so that all calculations are fully reproducible.

\section{Formalism: Conceptual Aspects}\label{Sect_Conceptual}

\begin{figure}
	\centering
	\includegraphics[width=1.05\linewidth]{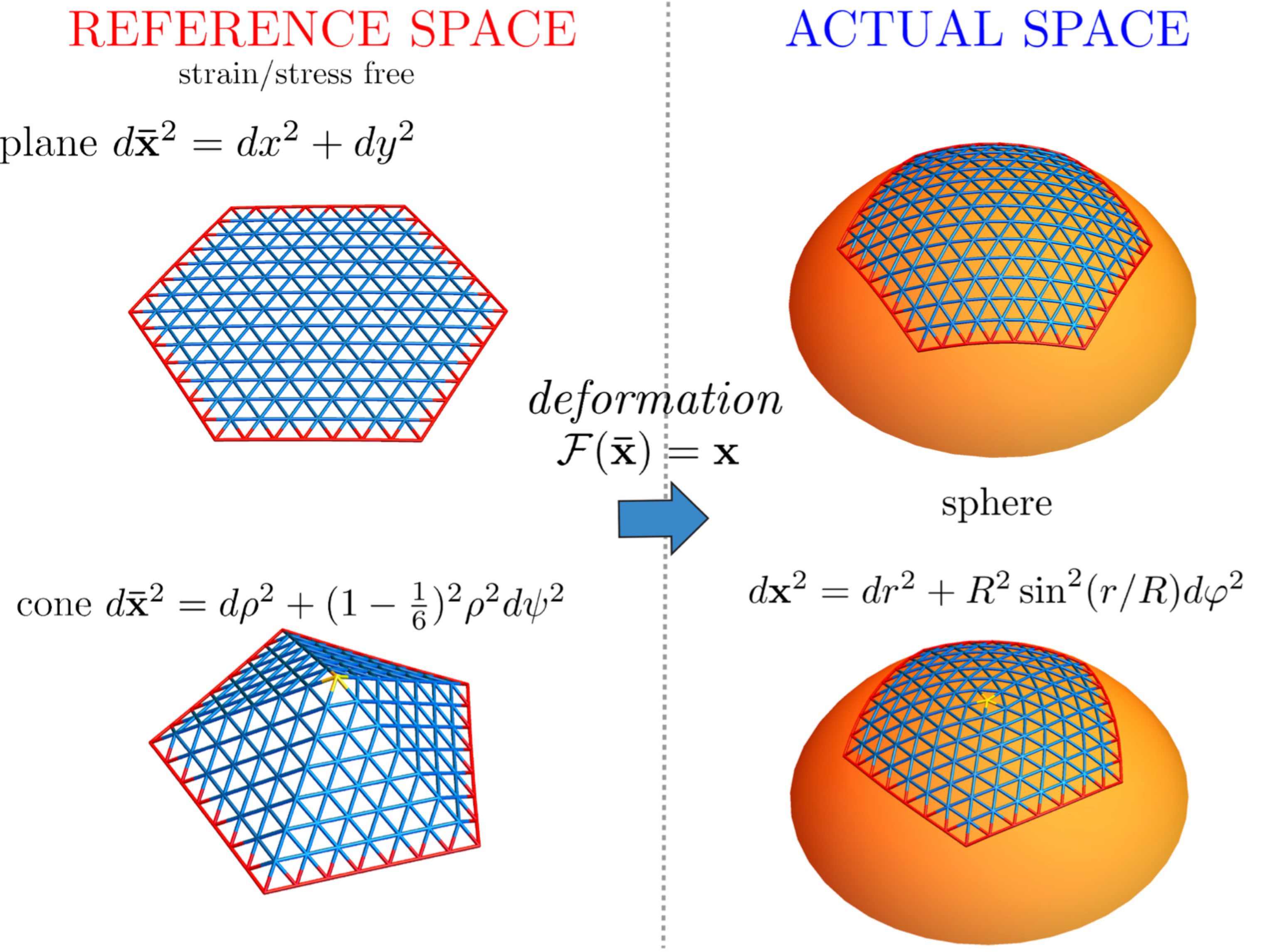}
	\caption{Example of reference/actual metric and space for a actual space consisting of a spherical cap. This problem is solved in Section~\ref{Sect_Results}.}
	\label{Fig:Ref_vsTarg}
\end{figure}

The basic quantities in elasticity theory are the displacements ${\bf u}({\bf \bar{x}})$ from a reference state ${\bf \bar{x}}$
\begin{equation}\label{Eq:EM_ref_state}
{\bf x} \equiv {\bf \bar{x}} + {\bf u}({\bf \bar{x}}) \ ,
\end{equation}
and the associated strain ($u_{\alpha\beta}$) and stress ($\sigma^{\alpha\beta}$) tensors, which are conjugated variables in the thermodynamic sense~\cite{LandauElasticityBook}. A definition of the strain tensor is given by comparing how a small vector in the {\em reference} (sometimes denoted as ``target'' \cite{Klein1116,Davidovitvh2019})  space $d{\bf \bar{x}}$ transforms after a mechanical deformation, represented by $d{\bf x}$:
\begin{eqnarray}\label{Eq:EM_ref_targz_metric}
d {\bf x}^2 & =& d {\bf \bar{x}}^2+ 2 u_{\alpha \beta}  d {\bf \bar{x}}^{\alpha} d {\bf \bar{x}}^{\beta}\ .
\end{eqnarray}
The physical interpretation of this equation is that two particles initially  apart by $d{\bf \bar{x}}$, after deformation become separated by $d{\bf x}$. This equation can be written as a function of two metrics, denoted as reference and actual hereon, as follows,
\begin{equation}\label{Eq:metric}
g_{\alpha \beta} = \bar{g}_{\alpha \beta} + 2 u_{\alpha \beta}.
\end{equation}
While the distances in the reference space are measured according to the metric $\bar{g}_{\alpha \beta}$, after deformation, which defines the {\em actual} space, distances and angles among physical particles change and are determined by the metric $g_{\alpha \beta}$, as illustrated in Fig.~\ref{Fig:Ref_vsTarg}. The strain tensor is the difference between actual and reference metrics.

The reference state is defined as a strain and stress free configuration, which is typically taken as ${\bf x}=(x,y,z)$ in 3D or ${\bf x}=(x,y,z=0)$ in 2D, which implies an euclidean reference metric
\begin{eqnarray}\label{Eq:EM_euclidean}
d {\bf \bar x}^2&=& dx^2+dy^2+dz^2 \mbox{ (3D) } \\
d {\bf \bar x}^2&=& dx^2+dy^2 \mbox{ (2D)}  \ .
\end{eqnarray}
Physically, the reference state maybe associated with a lattice where all nearest neighbors are at the same distance and form the same angle. In 2D we associate it with the triangular lattice, see Fig.~\ref{Fig:Ref_vsTarg}. Further below, we will show that the reference state is not unique, as a triangular lattice with topological defects such as disclinations and dislocations is also allowed. We mention, on passing, that in 3D a lattice where all nearest neighbors are at the same distance and form the same angle would consist of a tiling with regular tetrahedra, which is not possible~\cite{SadocBook1999} and leads to several consequences that have been discussed elsewhere~\cite{Nelson1983,Travesset2017b}.

Our goal in this paper is to develop a formalism to obtain the stress and strain in a curved surface. In particular, we focus on how an initially flat monolayer, whose reference state is given by ${\bf \bar x}$, consisting of a plane with additional defects, deforms into a given topography ${\vec r}({\bar{\bf x}})$ embedded in 3D space, as illustrated in Fig.~\ref{Fig:Ref_vsTarg}. Note that both the reference metric $d {\bf \bar x}^2$ and actual metric $d{\vec r}^2$ (which, in order to alleviate the notation will be denoted as $d{\bf x}^2$ in what it is, certainly, a blatant abuse of language) are known beforehand. We aim at finding the following transformation
\begin{equation}\label{Eq:EM_Form_application}
{\bf x} = {\cal F}({\bf \bar x}) \ ,
\end{equation}
which will be obtained by solving the equations of elasticity theory. How this transformation is related to the more familiar quantities in elasticity theory: the stress tensor $\sigma^{\alpha\beta}$, the Airy function ($\chi$)~\cite{NelsonBook2002}  etc.. will be discussed extensively later in the paper.

The problem of finding the transformation given in Eq.~\ref{Eq:EM_Form_application} is quite subtle because of the interplay of curvature, topology and defects such as disclinations or dislocations~\cite{NelsonBook2002, ChaikinBook2003}. Disclinations, for example, lead to long range effects that forbid many putative configurations; In a boundary free crystal, where the sum of all disclination charges is related to the Euler characteristic $\chi_E$ through the Gauss Bonnet theorem~\cite{Nakaharabook1990}
\begin{equation}\label{Eq:EM_Topo}
\sum_{i=1}^M s_i = \int d^2 {\bf x}\sqrt{g}K({\bf x}) = 2\pi \chi_E \ ,
\end{equation}
where $K({\bf x})$ is the Gaussian curvature, $g$ is the determinant of the surface metric and for a triangular lattice $s_i=\frac{\pi}{3} q_i$ ($q_i=\pm 1$). In case of a spherical surface, $\chi_E=2$ leading to the well known result that a spherical crystal has an excess of twelve $q_i=1$ disclinations (pentamers) in the absence of heptamers ($q_i=-1$).

Solutions to the theory of elasticity are obtained mostly within the Foppl Von Karman theory of elastic plates, which amounts to small displacements from equilibrium positions, an approach we denote as the Euler Framework (EF). A useful quantity to calculate the free energy and stress of a curved object is the Airy stress function.  For a crystal consisting of $M$ disclinations at positions ${\bf x}_i$ and with charge $s_i$, the equation for the Airy function is
\begin{equation}\label{Eq:EM_EF}
\frac{1}{Y} \Delta^2 \chi({\bf x}) = \sum_{i=1}^M s_i \delta({\bf x}_i - {\bf x}) - K({\bf x}) \ ,
\end{equation}
where $\Delta$ is the 2D Laplacian on a plane and $Y$ is the Young modulus~\cite{LandauElasticityBook,Seung1988}. Note that the Gaussian curvature of the surface acts as an external field. Relevant solutions to Eq.~\ref{Eq:EM_EF} are available for a buckled disclination or dislocation~\cite{Seung1988}, a spherical cap with and without a central disclination~\cite{Schneider2005, MorozovBruinsma2010} and also, for a spherical cap with an off-center disclination~\cite{Grason2010, Grason2012,Grason2015}. We emphasize again that the EF is exact in the limit of small curvature only. More precisely, if $r_m$ is the dimension of the crystal and $R$ some ``average'' curvature of the surface, the small curvature limit is defined by
\begin{equation}\label{Eq:EM_Param}
\alpha \equiv \frac{r_m}{R} = \theta_m << 1 \ .
\end{equation}
In a spherical cap (with constant curvature radius $R$),
a major problem arises as $\alpha \rightarrow \pi$, that is, as the spherical cap becomes a full sphere. Because within EF the solution of Eq.~\ref{Eq:EM_EF} is defined on a plane for a disk of area $A=\pi r^2_m$, the constraint Eq.~\ref{Eq:EM_Topo}
\begin{equation}\label{Eq:EM_Topo_viol}
\int d^2 {\bf x} K({\bf x}) = \int \frac{d^2{\bf x}}{R^2} =  \frac{A}{R^2} =  \pi \alpha^2 \neq 4\pi \ ,
\end{equation}
breaks down.

For a full sphere~\cite{Castelnovo2017}, the topological constraint Eq.~\ref{Eq:EM_Topo_viol} cannot be satisfied within EF. The failure to exactly satisfy a topological constraint is a serious conceptual problem that typically results in very significant computational errors. In 
Ref.~\cite{BowickMe2000,BowickMe2002,BowickMea2006} a generalization of Eq.~\ref{Eq:EM_EF}, which we denote as the Laplace Formalism (LF), was proposed
\begin{equation}\label{Eq:EM_LF}
\frac{1}{Y} \Delta^2_g \chi({\bf x}) = \frac{1}{\sqrt{g({\bf x})}}\sum_{i=1}^M s_i \delta({\bf x}_i - {\bf x}) - K({\bf x}) \ ,
\end{equation}
where the Laplacian $\Delta_g$ is computed with the actual metric, {\it i.e.}, on the curved surface. Now, for a full sphere, the topological constraint Eq.~\ref{Eq:EM_Topo} is satisfied identically. Although very successful and highly accurate in many applications~\cite{GiomiBowick2007}, the LF appears as an uncontrolled approximation: It is not obvious how to compute next orders so that eventually the exact solution will be recovered. Furthermore, for crystal with boundaries, like a crystal spanning a spherical cap, it is not immediately apparent what additional boundary conditions must be supplemented to Eq.~\ref{Eq:EM_LF}.

For the reasons exposed, neither the EF nor the LF are entirely satisfactory, despite their many successes. There is a clear need for a more rigorous formalism able to develop the LF as a systematic expansion and from which the EF appears as a low curvature expansion. A first insight on how to develop this formalism is provided by the fact that physical quantities (energies, stresses, strains, etc..) should be independent of surface parameterizations, that is, expressed in terms of geometric invariants, an approach pioneered by Kondo \cite{Kondo1955} in 1955 and Koiter as early as 1966~\cite{Koiter1966}. An elegant formulation with numerous new insights has been provided in Ref.~\cite{Efrati2009} and extended further in Ref.~\cite{MosheSharonKupferman2015}. In previous papers, see Ref.~\cite{Travesset2016a, LiMe2018} we have anticipated some aspects of the formalism fully elaborated here. 

Before dwelling into the actual formalism, it is worth describing the main ideas and concepts, which are very intuitive despite the significant amount of differential geometry~\cite{Nakaharabook1990} necessary for its rigorous development. As already discussed, both the actual metric  $g_{\mu \nu}({\bf x})$ and the reference ${\bar g}_{\mu \nu}({\bf \bar x})$ are known, what is therefore needed is the transformation Eq.~\ref{Eq:EM_Form_application} that enables to express the two metrics either as $g_{\mu \nu}({\bf \bar x})$ or ${\bar g}_{\mu \nu}({\bf x})$.

A simple counting of the number of variables helps understand the problem better. A general metric has three degrees of freedom $g_{11},g_{22},g_{12}$,
so in order to exactly map ${\bar g}_{\mu \nu}$ into $g_{\mu \nu}$ three functions are necessary. The solution of elasticity theory Eq.~\ref{Eq:EM_Form_application} provides only two of them as ${\cal F}$ is a 2D mapping. The third function is associated with the Gaussian curvature. If the curvature of the reference and actual metrics are not the same, a situation that is called {\em geometric frustration} or {\em metric incompatibility}, then it is not possible to make the two metrics ${\bar g}_{\mu \nu}$ and $g_{\mu \nu}$ coincide by Eq.~\ref{Eq:EM_Form_application}. Since the Gaussian curvature is a scalar invariant under reparameterizations, metric incompatibility, immediately leads to non-zero strains (and stresses), as obvious from Eq.~\ref{Eq:EM_ref_targz_metric}.

A few more clarifications are pertinent. First of all, as discussed above, the reference metric represents a strain and stress free configuration. Different from other descriptions, see Ref.~\cite{Efrati2013}, the reference metric does not have any residual strains/stresses, as it consists of patches of a flat metric joined by disclinations, where elasticity theory is not defined. The metric of a plane, representing a triangular lattice, is an example of a reference metric that can be embedded into actual space without any stresses. However, there are others: a cone with the appropriate aperture angle and $q=1,2,3$ disclination charge at its tip and $q=0$ (hexamers) everywhere else is also a stress and strain free configuration in the actual space. In the same way, one can consider a reference metric that contains an arbitrary number of defects, and hence, the associated curvature will be given by the disclination density $s({\bf \bar x})$
\begin{eqnarray}\label{Eq:EM_ref_curvature}
\bar{K}({\bf \bar x})&=&s({\bf \bar x})=\frac{1}{\sqrt{\bar g}}\sum_{j=1}^{M} s_j\delta({\bf \bar x}-{\bf \bar x}^j) \\\nonumber
    &=&\frac{1}{\sqrt{\bar g}}\left(\sum_{j=1}^{N_D}s_j\delta({\bf \bar x}-{\bf \bar x}^j)+
   \right. \\\nonumber &+&\left.  \sum_{i=1}^{N_d}\epsilon^{\alpha \beta}  b_{\alpha}^i\partial_{\mu}(\tensor{e}{_{\beta}^{\mu}}\delta({\bf \bar x}-{\bf \bar x}^i))\right)
\end{eqnarray}
where use has been made of vielbeins $\tensor{e}{_{\beta}^{\mu}}$, see appendix~\ref{App_Vielbeins}. The second equality follows by separating the $M$ disclinations as $N_D$ isolated disclinations and $N_d$ dislocations, that is, considering tightly bound disclinations as dipoles characterized by a Burgers vector ${\vec b}$. Only for a few cases, such as $N_D=0,N_d=0$ (plane), $N_D=1,N_d=0$ (cone) or $N_D=k, N_d=0$ (with 12 $\ge k \ge 2$, icosahedral sections), see also the limiting case $N_D=0, N_d=1$~\cite{Guven2013} as well as others, it is possible to embed explicit solutions in actual space such that $K=\bar{K}$ and therefore, they are strain and stress free. In this form, elasticity solutions amount to expressing a given metric $g_{\alpha \beta}$ as its optimal approximate in terms of ``quanta'' of disclinations of charge $\frac{\pi}{3}q$ and dislocations of Burgers vector ${\bf b}$. In fact, the geometric content of this ``quanta'' becomes even more explicit by noting that isolated disclinations are ``quanta'' of Gaussian curvature while dislocations are of geometrical torsion~\cite{BowickMed2001,Travesset2016a}.  

In this paper, we will not further discuss the role of dislocations, however, it is worth noting that it is possible to approximate any metric by Eq.~\ref{Eq:EM_ref_curvature} if $N_d\rightarrow \infty$, as demonstrated in 
Ref.~\cite{Travesset2016a}. This corresponds to the limit where Burgers vector ${\bf b}$ are infinitesimally small, {\it i.e.} mean field solutions, also discussed in Ref.~\cite{AzadiGrason2014,Azadi2016}. In this limit, the {\em Perfect Curvature Condition} (PCC)
\begin{equation}\label{Eq:EM_pcc}
K({\bf x})= s({\bf x})
\end{equation}
is satisfied. As pointed out in Ref.~\cite{IrvineVitelliChaikin2010}, it has the electrostatic analogy of a continuum of charge $K({\bf x})$ being represented by $N_D$ isolated charges and a continuum of polarization, i.e. $N_d \rightarrow \infty$ dipoles. More generally, the quantity 
\begin{equation}\label{Eq:EM_eta}
\eta({\bf x})=K({\bf x})-s({\bf x})
\end{equation}
is a measure of the geometric frustration or metric incompatibility. The PCC $\eta({\bf x})=0$ is the necessary and sufficient condition for a stress-strain free state to exist in actual space. We next develop these ideas in precise mathematical form.

\section{Formalism: Development}\label{Sect_Formalism}

\subsection{Exact Formulas}

As introduced previously, we will consider two metrics, $g_{\mu \nu}({\bf x})$ (actual metric) and ${\bar g}_{\mu \nu}({\bf x})$ (reference metric). The reference domain ${\cal B}_r$ represents the rest frame where the elastic energy is zero. The actual metric is defined over ${\cal B}_t$, which we denote as the {\em actual} domain. Consistent with our discussion in Sect.~\ref{Sect_Conceptual}, we will denote as ${\bf x}$ the actual coordinates and as ${\bf \bar x}$ the reference coordinates.  The solution of the problem is then to determine ${\cal F}$ in Eq.~\ref{Eq:EM_Form_application} (${\bf x} = {\cal F}({\bf \bar x})$).

The most general elastic free energy has the form
\begin{equation}\label{Eq:Form_Elastic}
  F = \frac{1}{2}\int_{{\cal B}} W\left(g({\bf x}), \bar{g}({\bf x})\right) d {\mbox Vol}_{g} \ .
\end{equation}
We now show that an appropriate choice of $W$ leads to the familiar expression for the elastic energy~\cite{LandauElasticityBook}, see also Ref.~\cite{MosheSharonKupferman2015}. If $Y$ is the Young modulus and $\nu_P$ is the Poisson ratio, the following quantities are defined
\begin{eqnarray}\label{Eq:form_A_def}
  A^{\alpha \beta \gamma \delta} & = & \frac{Y}{1-\nu^2_P}\left(\nu_P g^{\alpha \beta} g^{\gamma \delta} + (1-\nu_P) g^{\alpha \gamma} g^{\beta \delta}\right) \\\nonumber
  A_{\alpha \beta \gamma \delta} & = & \frac{1}{Y}\left( (1+\nu_P) g_{\alpha \gamma} g_{\beta \delta} - \nu_P g_{\alpha \beta} g_{\gamma \delta} \right)
\end{eqnarray}
in such a way that $A^{\alpha \beta \gamma \delta} A_{\gamma \delta \alpha^{\prime} \beta^{\prime}} = g^{\alpha}_{\alpha^{\prime}} g^{\beta}_{\beta^{\prime}}$. Then the functional $W\left(g({\bf x}), \bar{g}({\bf x})\right)$ is defined so that it reduces to the standard elastic energy for an isotropic medium, that is
\begin{equation}\label{Eq:form_W_def}
  W\left(g({\bf x}), \bar{g}({\bf x})\right) = A^{\alpha \beta \gamma \delta} u_{\alpha \beta} u_{\gamma \delta} \ ,
\end{equation}
where the strain tensor, see Eq.~\ref{Eq:EM_ref_targz_metric}, is
\begin{equation}\label{Eq:form_def_strain}
  2 u_{\alpha \beta}({\bf x}) = g_{\alpha \beta}({\bf x}) - {\bar g}_{\alpha \beta}({\bf x}) \ .
\end{equation}
Note that the free energy Eq.~\ref{Eq:Form_Elastic} is invariant under general reparameterizations. Working in the actual frame, the
metric $g_{\alpha \beta}({\bf x})$ is known, so we will derive the equilibrium equations in order to determine the reference metric ${\bar g}_{\mu \nu}({\bf x})$, which, expressed in the actual coordinates is not known. The stress tensor is given by
\begin{equation}\label{Eq:form_Stress_tensor}
  \sigma^{\alpha \beta} = \frac{1}{\sqrt{g}} \frac{\delta F}{\delta u_{\alpha \beta}} = A^{\alpha \beta \gamma \delta} u_{\gamma \delta} \ .
\end{equation}
Variations of Eq.~\ref{Eq:Form_Elastic} under reparameterizations ($\xi_{\beta}$) of the reference metric $\delta {\bar g}_{\alpha \beta} = -\bar{\nabla}_{\alpha}\xi_{\beta}-{\bar \nabla}_{\beta}\xi_{\alpha}$, leaving the actual metric invariant gives 
\begin{eqnarray}\label{Eq:form_variation}
  \delta F  & = & -\frac{1}{2}\int_{{\cal B}}d^2{\bf x} \sqrt{g} \sigma^{\alpha \beta} \delta {\bar g}_{\alpha \beta}
             =  \int_{{\cal B}}d^2{\bf x} \sqrt{g} \sigma^{\alpha \beta}\bar{\nabla}_{\alpha}\xi_{\beta}\nonumber \\
            & = &\int_{{\cal B}}d^2{\bf x} \left[\frac{\partial}{\partial x_{\alpha}} \left( \sqrt{g} \sigma^{\alpha \beta}\xi_{\beta} \right) \right. \nonumber\\
            &-&\sqrt{\bar g}\left.\bar{\nabla}_{\alpha} \left(\left(\frac{g}{\bar g}\right)^{1/2} \sigma^{\alpha \beta}\right) \xi_{\beta}\right]
\end{eqnarray}

The first term is a total derivative, and it can be converted to an integral along the boundary
\begin{eqnarray}\label{Eq:form_variation_boundary}
\int_{{\cal B}}d^2{\bf x} \frac{\partial}{\partial x_{\alpha}} \left( \sqrt{g} \sigma^{\alpha \beta} \xi_{\beta} \right) &=& \int_{\partial {\cal B}} dx^{\rho} \sqrt{g} \epsilon_{\rho \gamma} \sigma^{\gamma \beta} \xi_{\beta} \ .
\end{eqnarray}
Should the boundary contain a line tension term
\begin{equation}\label{Eq:form_line_tension}
F_l = \gamma \int_{\partial\cal B} ds \ ,
\end{equation}
then 
\begin{equation}\label{Eq:form_line_tension_var}
\delta F_l = -\gamma \int_{\partial \cal B} dx^{\mu}\nabla_{\mu} t^{\nu}\xi_{\nu} \ , 
\end{equation}
where $t^{\mu}$ is the unit tangent to the boundary. Taking into account the geometric formula
\begin{equation}\label{Eq:form_boundary_curvature}
t^{\mu}\nabla_{\mu} t^{\nu} = \frac{1}{r_{\cal B}} e\indices{_{\alpha}^{\nu}}n^{\alpha} \ ,
\end{equation}
with $r_{\cal B}$ the radius of curvature, $n^{\alpha}$ the normal and
$e\indices{_\alpha^\nu}$ are the vielbeins, see the appendix~\ref{App_Vielbeins}. The correct boundary condition is:
\begin{equation}\label{Eq:bound_cond}
n_{\gamma}\hat{\sigma}^{\gamma \nu}=-\frac{\gamma}{r_{\cal B}} n^{\nu} ,
\end{equation}
where $\hat{\sigma}^{\alpha \beta}=e\indices{^\alpha_\mu}e\indices{^\beta_\nu}\sigma^{\mu \nu}$, see appendix~\ref{App_Vielbeins} for the different expressions of the stress tensor and some additional details on the derivation of these formulas. This boundary condition reduces to the one derived for the EF in Ref.~\cite{MorozovBruinsma2010}.

From the definition of the covariant derivative, it is
\begin{equation}\label{Eq:form_covariant_derl}
  \nabla_{\alpha} \sigma^{\alpha \beta} = \frac{\partial \sigma^{\alpha \beta}}{\partial x_{\alpha}} +  \Gamma\indices*{^\alpha_{\alpha}_{\gamma}}\sigma^{\gamma \beta} + \Gamma\indices*{^\beta_{\alpha}_{\gamma}}\sigma^{\alpha \gamma} \ .
\end{equation}
Therefore, the equations determining equilibrium are
\begin{equation}\label{Eq:form_equations_bis}
\bar{\nabla}_{\alpha} \left(\left(\frac{g}{\bar g}\right)^{1/2} \sigma^{\alpha \beta}\right)
=
{\bar \nabla}_{\alpha}\sigma^{\alpha \beta} + \left(\Gamma\indices*{^\alpha_{\alpha}_{\gamma}} - {\bar \Gamma}\indices*{^\alpha_{\alpha}_{\gamma}}\right)\sigma^{\gamma \beta} = 0 \ ,
\end{equation}
which can also be written as
\begin{equation}\label{Eq:form_equations}
\nabla_{\alpha}\sigma^{\alpha \beta} + \left( {\bar \Gamma}\indices*{^\beta_{\alpha}_{\gamma}} - \Gamma\indices*{^\beta_{\alpha}_{\gamma}}\right)\sigma^{\alpha \gamma} = 0 \ ,
\end{equation}
derived first in Ref.~\cite{Efrati2009}. The appropriate boundary conditions as defined by Eq.~\ref{Eq:bound_cond}. Here, we have used the Christoffel symbols that are symmetric $\Gamma\indices*{^\beta_{\alpha}_{\gamma}} = \Gamma\indices*{^\beta_{\gamma}_{\alpha}}$. 

A general solution to Eq.~\ref{Eq:form_equations_bis} is given by the following ansatz~\cite{MosheSharonKupferman2015}
\begin{equation}\label{Eq:form_solution}
\sigma^{\alpha \beta} = \frac{1}{\sqrt{g}}\frac{1}{\sqrt{\bar g}} \epsilon^{\alpha \rho}\epsilon^{\beta \gamma} {\bar \nabla}_{\rho} {\bar \nabla}_{\gamma} \chi \ ,
\end{equation}
where $\epsilon^{12}=-\epsilon^{21}= 1$ and zero otherwise, and $\chi$ is the Airy function. Using the following identity,

\begin{equation}\label{Eq:App_Riemann_5}
  \frac{1}{g} \epsilon^{\alpha \rho}\epsilon^{\mu \nu} = g^{\alpha \mu} g^{\rho\nu} - g^{\alpha\nu} g^{\rho\mu}, \
\end{equation}
Eq.~\ref{Eq:form_solution} can be written as
\begin{equation}\label{Eq:form_solution_covariant}
\sigma^{\alpha \beta} = \left(\frac{\bar g}{g}\right)^{1/2} \left( {\bar g}^{\alpha \beta} {\bar g}^{\rho \gamma} - {\bar g}^{\alpha \gamma} {\bar g}^{\beta \rho} \right) {\bar \nabla}_{\rho} {\bar \nabla}_{\gamma} \chi.
\end{equation}
Using the formula $g^{\rho \gamma} \Gamma\indices*{^\nu_\rho_\gamma} = -\frac{1}{\sqrt{g}} \partial_{\gamma}(\sqrt{g}g^{\gamma \nu})$ and the fact that the covariant derivative of the metric is zero, {\it i.e.}, ${\bar \nabla}_{\alpha} {\bar g}_{\mu \nu} =0$, we find
\begin{equation}\label{Eq:form_actual_solution}
{\bar \nabla}_{\alpha} \sigma^{\alpha \beta} + \left(\Gamma\indices*{^\alpha_{\alpha}_{\gamma}} - {\bar \Gamma}\indices*{^\alpha_{\alpha}_{\gamma}}\right)\sigma^{\gamma \beta} = \frac{1}{\sqrt{g{\bar g}}} \epsilon^{\alpha \rho}\epsilon^{\beta \gamma} {\bar \nabla}_{\alpha} {\bar \nabla}_{\rho} {\bar \nabla}_{\gamma} \chi \ .
\end{equation}
The right hand side of the above equation can be expressed in terms of the Riemann tensor, see Eq.~\ref{Eq:App_Riemann_3}, as follows
\begin{eqnarray}\label{Eq:form_Riemmann_tensor}
\epsilon^{\alpha \rho}\epsilon^{\beta \gamma} {\bar \nabla}_{\alpha} {\bar \nabla}_{\rho} {\bar \nabla}_{\gamma} \chi & = &
\frac{1}{2}\epsilon^{\alpha \rho}\epsilon^{\beta \gamma} [{\bar \nabla}_{\alpha},{\bar \nabla}_{\rho}] {\bar \nabla}_{\gamma} \chi \nonumber \\
    &=& \frac{1}{2}\epsilon^{\alpha \rho}\epsilon^{\beta \gamma} {\bar R} \indices{^{\mu}_{\gamma}_{\alpha}_{\rho}} {\bar \nabla}_{\mu} \chi = 0 \ ,
\end{eqnarray}
where the last identity follows since the Riemann tensor of the reference metric is zero outside the defect cores, that is, almost everywhere, see Eq.~\ref{Eq:EM_ref_curvature}. Thus, Eq.~\ref{Eq:form_solution} provides a general solution of Eq.~\ref{Eq:form_equations_bis} in terms of the Airy function.

Substituting the solution of Eq.~\ref{Eq:form_solution} into the definition of the strain Eq.~\ref{Eq:form_def_strain} gives,
\begin{equation}\label{Eq:form_elastic_equation}
  \frac{1}{\sqrt{g}}\frac{1}{\sqrt{\bar g}} \epsilon^{\alpha \rho}\epsilon^{\beta \gamma} {\bar \nabla}_{\rho} {\bar \nabla}_{\gamma} \chi = \frac{1}{2} A^{\alpha \beta \gamma \delta} \left(g_{\gamma \delta} - {\bar g}_{\gamma \delta} \right)
\end{equation}
or
\begin{eqnarray}\label{Eq:form_elastic_equation_m}
  {\bar g}_{\alpha \beta} &=& g_{\alpha \beta} - \frac{2}{\sqrt{g\bar g}}A_{\mu \lambda \alpha \beta}\epsilon^{\mu \rho}\epsilon^{\lambda \gamma} {\bar \nabla}_{\rho} {\bar \nabla}_{\gamma} \chi \\\nonumber
  {\bar g}_{\alpha \beta} &=& g_{\alpha \beta} - \frac{2}{Y}\left(\frac{g}{\bar g}\right)^{1/2}\left[g_{\alpha \beta} g^{\rho\gamma} - (1+\nu_P)g^{\gamma}_{\alpha} g^{\rho}_{\beta}\right] {\bar \nabla}_{\rho} {\bar \nabla}_{\gamma} \chi
\end{eqnarray}
Thus ${\bar g}_{\mu \nu}(\chi({\bf x}))$ can be obtained from above equation. Note however, that among all possible functions $\chi$, there is only a unique family that has the right curvature ${\bar K}$, so the equation above needs to be supplemented with the  additional constraint
\begin{equation}\label{Eq:form_elastic_equation_m_reference}
2{\bar K} = {\bar R} = {\bar g}^{\mu \nu} {\bar R}_{\mu \nu} = {\bar g}^{\mu \nu} {\bar R}\indices{^\rho_\mu_\rho_\nu} = 0 \ ,
\end{equation}
which uniquely determines the family of solutions $\chi$. Here ${\bar K}= s({\bf x})$ is the Gaussian curvature, ${\bar R}$ the scalar curvature, ${\bar R}_{\mu \nu}$ the Ricci tensor and ${\bar R}\indices{^\rho_\mu_\gamma_\nu}$ the Riemann tensor. That is, the solution consists among all possible functions of $\chi$, to select the one that makes ${\bar g}_{\mu \nu}$ a quasi-flat metric. In general, such solution is complicated as ${\bar g}_{\mu \nu}$ appears on both sides of the equation, and the rhs includes its derivatives. The mapping ${\bf x}={\cal F}({\bar {\bf x}})$ is obtained as the three dimensional vector field whose metric is $g$. A concrete example is discussed further below.

Using Eqs.~\ref{Eq:form_W_def}-\ref{Eq:form_Stress_tensor} and \ref{Eq:form_elastic_equation_m}, the expression for the elastic energy (Eq.~\ref{Eq:Form_Elastic}) without any approximations is,
\begin{eqnarray}\label{Eq:form_elastic}
  F & = & \frac{1}{2}\int_{{\cal B}} \sigma^{\alpha \beta} A_{\alpha \beta \rho \sigma} \sigma^{\rho \sigma} d {\mbox Vol}_{g} \\\nonumber
  &=& \frac{1}{2Y} \int_{{\cal B}} d {\mbox Vol}_{g} \frac{g}{{\bar g}}\left((1+\nu_p)g^{\alpha \rho} g^{\beta \sigma}-\nu_p g^{\alpha \beta} g^{\rho \sigma} \right) \times \nonumber \\
  && \times {\bar \nabla}_{\alpha}{\bar \nabla}_{\beta} \chi {\bar \nabla}_{\rho}{\bar \nabla}_{\sigma} \chi 
\end{eqnarray}
note that up to this point all formulas are exact. We now discuss some common approximations. 

\subsection{Incompatibility metric approximation}

\subsubsection{Actual frame}

Since the actual metric $g_{\mu\nu}({\bf x})$ is known, the goal is to compute the reference metric $\bar{g}_{\mu\nu}({\bf x})$, and from there, one can obtain the transformation Eq.~\ref{Eq:EM_Form_application}. If one assumes that $\eta$, see Eq.~\ref{Eq:EM_eta}, is somehow small, the Airy function and the metric are: 
\begin{eqnarray}\label{Eq:form_expansion}
   \chi &=&  \chi^{(1)} + \chi^{(2)} + \cdots \\
  {\bar g} &=& g + {g}^{(1)} + {g}^{(2)} + \cdots \ ,
\end{eqnarray}
where each term contains increasing powers of $\eta$. Obviously the Airy function is at least, linear with $\eta$, as for $\eta=0$, $\chi=0$ and $g=\bar{g}$. Plugging this expansion into the Airy equation~\ref{Eq:form_elastic_equation_m} provides the explicit orders in the expansion. The first order is
\begin{equation}\label{Eq:form_orders}
g\indices*{_\alpha_\beta^{(1)}} = -\frac{2}{Y}\left( g_{\alpha \beta} \Delta \chi^{(1)} - (1+\nu_P)\nabla_{\alpha}\nabla_{\beta} \chi^{(1)} \right) \ ,
\end{equation}
where $\Delta = g^{\alpha \beta}\nabla_{\alpha}\nabla_{\beta} = \frac{1}{\sqrt{g}} \partial_{\alpha} (g^{\alpha\beta} \sqrt{g} \partial_{\beta})$ is the Laplace-Beltrami operator. Higher orders are discussed in the appendix~\ref{App_IM}. The goal is now to derive an explicit equation for $\chi^{(i)}$, as discussed below.

\subsubsection{First order expressions for energy and stress: actual frame}

With the metric expressed linearly in terms of the Airy function, the next step is to enforce the constraint Eq.~\ref{Eq:form_elastic_equation_m_reference}. For this purpose, it is necessary to compute the scalar curvature. This calculation is relegated to appendix~\ref{App_IM}, and gives
\begin{eqnarray}\label{Eq:form_linear_scalar}
{\bar K} &=& K +\frac{1}{Y} \left(\Delta^2 \chi^{(1)} + \right. \\\nonumber
&+& \left. 2K \Delta\chi^{(1)}+(1+\nu_p) g^{\mu\lambda} \nabla_{\mu}K \nabla_{\lambda} \chi^{(1)}\right) \ .
\end{eqnarray}
In addition to the square of Laplacian in the above equation there are additional terms that will be explored further below. The stress tensor within this order is
\begin{equation}\label{Eq:form_linear_stress}
  \sigma^{\alpha \beta} = g^{\alpha \beta} \Delta \chi^{(1)} - g^{\alpha\mu}g^{\beta\nu} \nabla_{\mu}\nabla_{\nu} \chi^{(1)} \ ,
\end{equation}
and the energy
\begin{eqnarray}\label{Eq:form_linear_free_energy}
  F & = & \frac{1}{2Y} \int d^2u \sqrt{g} \left[ (\Delta \chi^{(1)})^2 +\right. \nonumber \\
   &+&\left. \frac{(1+\nu_P)}{g} \epsilon^{\alpha \sigma}\epsilon^{\rho \beta} \nabla_{\alpha}\nabla_{\beta} \chi^{(1)} \nabla_{\rho}\nabla_{\sigma} \chi^{(1)}\right] .
\end{eqnarray}
As elaborated in appendix~\ref{App_Energy_Target}, may be expressed as
\begin{eqnarray}\label{Eq:form_linear_free_energy_final}
F & = & \frac{1}{2Y} \int d^2u \sqrt{g} (\Delta \chi^{(1)})^2 -\\ \nonumber
&-& \frac{1+\nu_p}{2Y}\int d^2u\sqrt{g} K g^{\alpha \beta}\nabla_{\alpha} \chi^{(1)}\nabla_{\beta} \chi^{(1)} - \\ \nonumber
&-&\frac{1+\nu_p}{2Y}\oint dx^{\rho} \sqrt{g}\epsilon_{\rho \alpha} \sigma^{\alpha\beta} \nabla_{\beta}  \chi^{(1)}.
\end{eqnarray}

A variation on the previous expansion consists in dropping the cross terms 
involving $K\chi$ in Eq.~\ref{Eq:form_linear_scalar}. The resulting equations are
\begin{equation}\label{Eq:form_linear_LF}
{\bar K} = K +\frac{1}{Y}\Delta^2 \chi^{(1)} \ ,
\end{equation}
with corresponding energy
\begin{eqnarray}\label{Eq:form_linear_energy_LF}
F & = & \frac{1}{2Y} \int d^2u \sqrt{g} (\Delta \chi^{(1)})^2  - \\ \nonumber
&-&\frac{1+\nu_p}{2Y}\oint dx^{\rho} \sqrt{g}\epsilon_{\rho \alpha} \sigma^{\alpha\beta} \nabla_{\beta}  \chi^{(1)} \ ,
\end{eqnarray}
which we recognize as the LF discussed in Sect.~\ref{Sect_Conceptual}. Note that in the absence of line tension or external stress, the boundary conditions determine that the second term vanishes identically.  Hereon, we will refer the approximation Eq.~\ref{Eq:form_linear_scalar} as the Incompatibility Framework (IF) in order to differentiate it from the LF.

\subsubsection{Reference frame}

The expansion for the metric and the Airy function is
\begin{eqnarray}\label{Eq:form_expansion_ref}
   \chi &=&  \chi^{(I)} + \chi^{(II)} + \cdots \\
   g &=& {\bar g} + {\bar g}^{(I)} +  {\bar g}^{(II)} + \cdots
\end{eqnarray}
Similarly as in the actual approximation Eq.~\ref{Eq:form_orders}, the first order is
\begin{equation}\label{Eq:form_orders_ref}
{\bar g}\indices*{_\alpha_\beta^{(I)}} = \frac{2}{Y}\left( {\bar g}_{\alpha \beta} {\bar \Delta} \chi^{(I)} - (1+\nu_P){\bar \nabla}_{\alpha}{\bar \nabla}_{\beta} \chi^{(I)} \right) \ ,
\end{equation}
with ${\bar \Delta}$ being the Laplace-Beltrami operator of the reference metric. Higher orders are discussed in the appendix~\ref{App_IM}.

\subsubsection{First order expressions for energy and stress: reference frame}

The formulas derived in the previous case automatically translate into the reference frame by replacing $ g_{\alpha \beta} \leftrightarrow {\bar g}_{\alpha \beta}$ and $\chi^{(1)} \rightarrow - \chi^{(I)}$, leading to

\begin{eqnarray}\label{Eq:form_linear_scalar_ref}
K &=& {\bar K} - \frac{1}{Y} \left({\bar \Delta}^2 \chi^{(I)} + \right. \\\nonumber
&+& \left. 2{\bar K} {\bar \Delta}\chi^{(I)}+(1+\nu_p) {\bar g}^{\mu\lambda} {\bar \nabla}_{\mu} {\bar K} {\bar \nabla}_{\lambda} \chi^{(I)}\right)
\end{eqnarray}
The stress tensor within this order is
\begin{equation}\label{Eq:form_linear_stress_ref}
  \sigma^{\alpha \beta} = {\bar g}^{\alpha \beta} {\bar \Delta} \chi^{(I)} - {\bar g}^{\alpha\mu}{\bar g}^{\beta\nu}  {\bar \nabla}_{\mu}{\bar \nabla}_{\nu} \chi^{(I)} \ ,
\end{equation}
and the energy
\begin{eqnarray}\label{Eq:form_linear_free_energy_ref}
  F & = & \frac{1}{2Y} \int d^2u \sqrt{\bar g} \left[ ({\bar \Delta} \chi^{(I)})^2 +\right. \nonumber \\
   &+&\left. \frac{(1+\nu_P)}{\bar g} \epsilon^{\alpha \sigma}\epsilon^{\rho \beta} {\bar \nabla}_{\alpha}{\bar \nabla}_{\beta} \chi^{(I)} {\bar \nabla_{\rho}}{\bar \nabla}_{\sigma} \chi^{(I)}\right] \ .
\end{eqnarray}
Given the assumptions about the reference metric, see Eq.~\ref{Eq:EM_ref_curvature}, the above equations simplify to
\begin{eqnarray}\label{Eq:form_linear_ref_sim}
\frac{1}{Y}{\bar \Delta}^2 \chi^{(I)} &=& {\bar K}-K 
\end{eqnarray}
and energy 
\begin{eqnarray}\label{Eq:form_linear_free_energy_sim}
F & = & \frac{1}{2Y} \int d^2u \sqrt{\bar g} ({\bar \Delta} \chi^{(I)})^2 
\end{eqnarray}
where $\bar{\Delta}$ is the Laplacian on the plane. Thus, the reference frame expansion coincides with the EF discussed in Sect.~\ref{Sect_Conceptual}. The singular terms in Eq.~\ref{Eq:EM_ref_curvature} can be dropped from the second term in Eq.~\ref{Eq:form_linear_free_energy_ref} as they only contribute within the defect cores. These contributions are accounted by an empirical core energy term $E_{core}$ as linear elasticity breaks down.

\section{Results}\label{Sect_Results}

As a concrete example, we will solve the case of a crystal on a sphere of radius $R$, as illustrated in Fig.~\ref{Fig:Ref_vsTarg}. The extent of the crystal is parameterized by its aperture angle $\theta_M$. This problem has been described previously within the EF by Schneider and Gommper\cite{Schneider2005} as well as Morozov and Bruinsma~\cite{MorozovBruinsma2010} as well as Grason~\cite{Grason2012}. In the current notation, the Gaussian curvature is $K=\frac{1}{R^2}$ and ${\bar K}$ the disclination density ${\bar K} = s({\bf r})$.
The reference frame metric is Euclidean and is defined over a disk of radius $\rho_0$ by
\begin{equation}\label{Eq:Results_reference}
  ds^2 = d\rho^2 + \rho^2\left(1-\frac{s}{2\pi}\right)^2 d\psi^2 \equiv {\bar g}_{\mu \nu} d{\bar x}^{\mu} d{\bar x}^{\nu} \ .
\end{equation}
The case $s=\frac{\pi}{3} q_i$ corresponds to a disclination of positive charge placed at the center of the disk. The actual metric is
\begin{equation}\label{Eq:Results_target}
  ds^2 = dr^2 + R^2\sin^2(r/R) d\varphi^2 \equiv g_{\mu \nu} dx^{\mu} dx^{\nu} \ .
\end{equation}
The problem then consists in finding the function ${\cal F}$ such that
\begin{equation}\label{Eq:Results_functionF}
  x^{\mu} = {\cal F}({\bar x}^{\mu}) \ ,
\end{equation}
where $x^{\mu}=(r, \varphi)$ and $\bar{x}^{\mu}=(\rho,\psi)$.
We will investigate symmetric solutions where $\psi=\varphi$
\begin{equation}\label{Eq:Results_functionF_symmetric}
  r \equiv r(\rho) = F(\rho)  \ ,
\end{equation}
so that the problem becomes one dimensional.

\subsection{Exact Solution}

We will discuss symmetric solutions defined by Eq.~\ref{Eq:Results_functionF_symmetric} and we will calculate $\rho(r)$. The reference metric is
\begin{equation}\label{Eq:Results_exact_ref_metric}
ds^2 = d\rho^2+\rho^2 d\psi^2 \equiv (\rho^{\prime}(r))^2 dr^2+w^2\rho^2(r) d\varphi^2 
\end{equation}
where $\rho^{\prime}=d\rho/dr$, $w\equiv 1-\frac{s}{2\pi}$ and the reference metric is expressed in actual coordinates. The non-zero Christoffel symbols are: 
\begin{equation}\label{Eq:Results_Christoffel}
\begin{array}{c c c c}
\mbox{symbol} & \Gamma\indices*{^r_{r}_{r}} & \Gamma\indices*{^r_{\varphi}_{\varphi}} & \Gamma\indices*{^{\varphi}_{\varphi}_{r}} \\
\mbox{reference} & \frac{\rho^{\prime\prime}(r)}{\rho^{\prime}(r)} & -w^2\frac{\rho(r)}{\rho^{\prime}(r)} & \frac{\rho^{\prime}(r)}{\rho(r)}\\
\mbox{actual}  & 0 & -R\sin(r/R)\cos(r/R) & \frac{\cot(r/R)}{R}
\end{array}
\end{equation}
The components of the stress tensor Eq.~\ref{Eq:form_Stress_tensor} is the difference between the actual and reference metric, that is
\begin{eqnarray}\label{Eq:Results_stress}
\sigma^{rr} &=&\frac{Y}{2(1-\nu^2_p)}\left[1-\rho^{\prime}(r)^2+\nu_p\left(1-\left(\frac{w\rho(r)}{R\sin(r/R)}\right)^2\right)\right] \nonumber\\
\sigma^{r \varphi}&=&0 \\ \nonumber
\sigma^{\varphi \varphi}&=& \frac{Y}{2(1-\nu^2_p)R^2\sin^2(r/R)}\times \\ \nonumber
   &\times& \left[ 1-\left(\frac{w\rho(r)}{R\sin(r/R)}\right)^2+
\nu_p(1-\rho^{\prime}(r)^2)\right] \ .
\end{eqnarray}
Inserting Eq.~\ref{Eq:Results_Christoffel} into Eq.~\ref{Eq:form_equations_bis} we obtain
\begin{eqnarray}\label{Eq:Results_eq_diff}
\frac{d \sigma^{rr}}{dr}+\Gamma\indices*{^\varphi_{\varphi}_{r}}\sigma^{rr}+ \bar{\Gamma}\indices*{^r_{r}_{r}}\sigma^{rr}+\bar{\Gamma}\indices*{^r_{\varphi}_{\varphi}}\sigma^{\varphi \varphi}  &=&0,
\end{eqnarray}
which becomes
\begin{equation}\label{Eq:stress11}
\frac{d \sigma^{rr}}{dr}+\left(\frac{\cot\left(\frac{r}{R}\right)}{R}+\frac{\rho^{\prime\prime}(r)}{\rho^{\prime}(r)}\right)\sigma^{rr}-\frac{w^2\rho(r)}{\rho^{\prime}(r)}\sigma^{\varphi \varphi}  = 0.
\end{equation}

Introducing Eq.~\ref{Eq:Results_stress} into Eq.~\ref{Eq:stress11} yields a nonlinear ordinary differential equation for $\rho(r)$
\begin{align}\label{Eq:Exact_eq_rho}
&\frac{2v w^2}{R^2\sin(\frac{r}{R})^2} \rho(r)^2\left(\frac{ \cot(\frac{r}{R})}{R}-\frac{\rho'(r)}{\rho
(r)}\right)-2 \rho'(r) \rho''(r) \nonumber\\
&+\left(\frac{\cot(\frac{r}{R})}{R}+\frac{\rho''(r)}{\rho'(r)}\right) \nonumber\\
&\hspace{1cm}\times \left[1-\rho'(r)^2+v\left(1-\frac{w^2 \rho(r)^2}{R^2\sin(\frac{r}{R})^2}\right)\right] \nonumber\\
&-w^2 \frac{\rho(r)}{\rho'(r)}\frac{1}{R^2\sin(\frac{r}{R})^2} \nonumber\\
&\hspace{1cm}\times\left[1-\frac{w^2\rho(r)^2}{R^2\sin(\frac{r}{R})^2}+v-v
\rho'(r)^2\right]=0
\end{align}
with boundary conditions $\rho(0)=0$ and 
$\sigma^{rr}(\theta_m R)=\frac{Y}{1-\nu^2_p}\left[1-\rho '(\theta_m  R)^2+v\left(1-\frac{w^2 \rho(\theta_m R)^2}{R^2\sin(\theta_m)^2}\right)\right]=0$.
Although within this formalism the Airy function is not necessary to calculate the stress, its actual form is valuable as a comparison with its approximations. It is given as:
\begin{eqnarray}\label{Eq:Results_Airy_exact}
\sigma^{r r} &= & 
\frac{1}{R\sin(r/R)w\rho(r)\rho^{\prime}(r)} \bar{\nabla}^2_{\varphi} \chi 
=\frac{w}{R\sin(r/R)\rho^{\prime}(r)^2}\frac{d\chi}{dr} \nonumber\\
\sigma^{\varphi \varphi}&=&
\frac{1}{R\sin(r/R)w\rho^{\prime}(r)\rho(r)}\left(
\frac{d^2 \chi}{dr^2}-\frac{\rho^{\prime\prime}(r)}{\rho^{\prime}(r)}\frac{d\chi}{dr}\right)
\end{eqnarray}
where $\sigma^{r \varphi}=0$ is satisfied identically. Note that only one of the equations needs to be satisfied, as the other becomes then an identity.

\subsection{Incompatibility metric approximation solutions}

\subsubsection{Reference frame}

The equations describing the Airy function for a disclination of charge $s$ in the reference frame have been described above, namely
\begin{equation}\label{Eq:Results_SN}
{\bar \Delta}^2 \chi^{(I)} + Y ( K - s({\bf r})) = 0  \ .
\end{equation}
The solution can be read directly from Ref.~\cite{MorozovBruinsma2010}, and it is given by
\begin{equation}\label{Eq:Results_Chi_ref}
\chi^{(I)}(\rho) = \frac{Y}{64R^2}\left(2\rho_0^2\rho^2-\rho^4\right) + \frac{Y s}{8\pi} \rho^2\left(\log(\rho/\rho_0)-\frac{1}{2}\right) \ ,
\end{equation}
where $\rho_0 = R \theta_m$ is the radius of the crystal. This is a double expansion in the small parameters $\rho_0^2/R^2$ and $s/(2\pi)$.

Substitution of Eq.~\ref{Eq:Results_Chi_ref} into Eq.~\ref{Eq:form_orders_ref} gives
\begin{eqnarray}
{\bar g}_{rr}^{(I)}&=&\frac{1}{8 R^2} (\rho_0^2 - \rho^2 + \nu_p (3 \rho^2 - \rho_0^2))-\frac{s}{2\pi} \nu_p \nonumber\\
&&+ \frac{s}{2\pi} (1 - \nu_p) \log(\frac{\rho}{\rho_0}), \nonumber\\
{\bar g}_{\phi\phi}^{(I)}&=&w^2 \rho^2 \left(\frac{1}{8 R^2} (\rho_0^2 - 3 \rho^2 + \nu_p (\rho^2 - \rho_0^2 )) \right. \nonumber\\
&&\left. + \frac{s}{2\pi} + \frac{s}{2\pi} (1-\nu_p) \log(\frac{\rho}{\rho_0})\right) \ .
\end{eqnarray}
The actual frame metric becomes
\begin{eqnarray}
g_{rr}&=&{\bar g}+{\bar g}_{rr}^{(I)} \nonumber\\
&=&1+\frac{1}{8 R^2} (\rho_0^2 - \rho^2 + \nu_p (3 \rho^2 - \rho_0^2)) \nonumber\\
&&-\frac{s}{2\pi} \nu_p + \frac{s}{2\pi} (1 - \nu_p) \log(\frac{\rho}{\rho_0}) \nonumber\\
&\equiv& r'(\rho)^2, \nonumber\\
g_{\phi\phi}&=&{\bar g}_{\phi\phi}+{\bar g}_{\phi\phi}^{(I)}\nonumber\\
&=&w^2 \rho^2+w^2 \rho^2 \left(\frac{1}{8 R^2} (\rho_0^2 - 3 \rho^2 + \nu_p (\rho^2 - \rho_0^2 )) \right. \nonumber\\
&&\left. + \frac{s}{2\pi} + \frac{s}{2\pi} (1-\nu_p) \log(\frac{\rho}{\rho_0})\right) \nonumber\\
&\equiv& \sin^2(r(\rho)) \ .
\end{eqnarray}
Using the transformation properties of $g(\bar{x})_{\mu \nu}$ in terms of ${\cal F}$ in Eq.~\ref{Eq:EM_Form_application}, we obtain
\begin{eqnarray}\label{Eq:Results_r_mapping}
r(\rho)
&=&\rho \left(1+\frac{1}{16 R^2} (\rho_0^2 - \frac{\rho^2}{3} + \nu_p (\rho^2 - \rho_0^2 )) \right. \nonumber\\
&&\left. - \frac{s}{4\pi} + \frac{s}{4\pi} (1-\nu_p) \log(\frac{\rho}{\rho_0})\right) \ ,
\end{eqnarray}
which is inverted to give the complete solution,
\begin{eqnarray}\label{Eq:Results_rho_mapping}
\rho(r)
&=&r \left(1-\frac{1}{16 R^2} ((\theta_m R)^2 - \frac{r^2}{3} + \nu_p (r^2 - (\theta_m R)^2 )) \right. \nonumber\\
&&\left. + \frac{s}{4\pi} - \frac{s}{4\pi} (1-\nu_p) \log(\frac{r}{\theta_m R})\right) \ .
\end{eqnarray}
The stresses are then found using Eq.~\ref{Eq:form_linear_stress_ref}
\begin{eqnarray}\label{Eq:stress_ref_EF}
\sigma^{\rho\rho}&=& \frac{Y}{16R^2} (\rho_0^2 - \rho^2) + \frac{Ys}{4 \pi} \log(\frac{\rho}{\rho_0})
\nonumber\\
\rho^2\sigma^{\psi\psi}&=&
\frac{Y}{16R^2} (\rho_0^2 - 3 \rho^2) +
\frac{Ys}{4 \pi} \left(1 + \log(\frac{\rho}{\rho_0}) \right)
\nonumber\\
\end{eqnarray}
and the free energy from Eq.~\ref{Eq:form_linear_free_energy_sim} becomes,
\begin{eqnarray}\label{Eq:FE_ref}
\frac{F}{\pi \rho_0^2 Y}&=&\frac{\theta_m^4}{384} + \frac{1}{32} (\frac{s^2}{\pi^2} - \frac{s}{2 \pi} \theta_m^2) \\ \nonumber
\frac{F}{\mbox{Area}\cdot Y} &=& \frac{\theta_m^4}{1536}+\frac{1}{32}\left(\frac{s}{\pi}-\frac{\theta_m^2}{4}\right)^2.
\end{eqnarray}
The limit $\theta_m \rightarrow 0$ (flat limit) agrees with previous results~\cite{Seung1988}.


\subsubsection{Actual frame}

With the assumptions that  $\psi=\varphi$, the actual metric becomes 
\begin{equation}\label{Eq:Results_target_metric}
ds^2 = ({\cal F}^{\prime}(\rho))^2 d\rho^2 + \sin^2({\cal F}(\rho)) d\psi^2
\end{equation}
The equations for the Airy function are either Eq.~\ref{Eq:form_linear_scalar} (IF) or Eq.~\ref{Eq:form_linear_LF} (LF), namely
\begin{eqnarray}\label{Eq:Results_target_chi}
\Delta^2 \chi^{(1)}_{IF} + \frac{2}{R^2}\Delta\chi^{(1)}_{IF} &=& s({\bf x}) - \frac{1}{R^2} \mbox{ (IF)} \\ \nonumber
\Delta^2 \chi^{(1)}_{LF}  &=& s({\bf x}) - \frac{1}{R^2} \mbox{  (LF)}
\end{eqnarray}
where $s({\bf x})$ is the disclination density. 

The solutions to Eq.~\ref{Eq:Results_target_chi} is
\small
\begin{eqnarray}\label{Eq:Results_Chi_tar_cur}
\chi^{(1)}_{IF}&&(r) /(Y R^2)=\log(\cos(\frac{r}{2R})) -\log(\cos(\frac{\theta_m}2)) \nonumber\\
&&-\frac{1}{2} \cos(\frac{r}{R}) \csc(\theta_m) \tan(\frac{\theta_m}2)+ \frac{1}{2} \cot(\theta_m) \tan(\frac{\theta_m}2) \nonumber\\
&&+\frac{s}{2\pi}\left[ \sin^2(\frac{r}{2R})\log(\frac{\tan(\frac{r}{2R})}{\tan(\frac{\theta_m}2)})  \right.\nonumber\\
&&\hspace{1cm}\left.-\frac{1}{2}\sin^2(\frac{r}{2R})  \sec^2(\frac{\theta_m}2) + \frac{1}{2} \tan^2(\frac{\theta_m}2)\right]
\end{eqnarray}
\normalsize
and also
\small
\begin{eqnarray}\label{Eq:Results_Chi_tar}
&&\chi^{(1)}_{LF}(r) /(Y R^2)= 
\text{Li}_2(\sin^2(\frac{r}{2R})) - \text{Li}_2(\sin^2(\frac{\theta_m}{2})) \nonumber\\
&&-\cot^2(\frac{\theta_m}{2}) \log(\frac{1 + \tan^2(\frac{r}{2R})}{1 + \tan^2(\frac{\theta_m}{2})}) \log(1 + \tan^2(\frac{\theta_m}{2})) \nonumber\\
&&+\frac{s}{2\pi}
\left[\text{Li}_2(-\tan^2(\frac{r}{2R})) - \text{Li}_2(-\tan^2(\frac{\theta_m}{2})) \right. \nonumber\\
&&\left.+ \log(\tan(\frac{r}{2R})) \log(1 + \tan^2(\frac{r}{2R})) \right.\nonumber\\
&&\left.- \log(\tan(\frac{\theta_m}{2})) \log(1 + \tan^2(\frac{\theta_m}{2})) \right.\nonumber\\
&&\left. +2 \log(\cos(\frac{r}{2R})) \left(\cot^2(\frac{\theta_m}{2}) \log(\cos(\frac{\theta_m}{2})) + \log(\sin(\frac{\theta_m}{2}))\right)\right. \nonumber\\ 
&&\left. - 2 \log(\cos(\frac{\theta_m}{2})) \left(\cot^2(\frac{\theta_m}{2}) \log(\cos(\frac{\theta_m}{2})) + \log(\sin(\frac{\theta_m}{2}))\right) \right] \ , \nonumber\\
\end{eqnarray}
\normalsize
with $Li_2$ the dilogarithmic function. It is relevant at this point to compare the Airy function in actual space with the one in reference space; the difference between both gives an idea of the errors involved in the coresponding approximations. Using Eq.~\ref{Eq:Results_Chi_ref} by expanding Eq.~\ref{Eq:Results_Chi_tar} to the next orders gives
\begin{eqnarray}\label{Eq:Results_Chi_tar_exp}
\chi^{(1)}_{IF}(x)&& /(Y R^2)=\nonumber\\
&&-\frac{1}{64} (x^2 - \theta_m^2)^2 + \frac{s}{16 \pi} (\theta_m^2 - x^2 + 2 x^2 \log(\frac{x}{\theta_m})) \nonumber\\
&&- \frac{1}{384} (\theta_m^6 - 2 x^2 \theta_m^4 + x^4 \theta_m^2) \nonumber\\
&&+ \frac{s}{192 \pi} (3 x^4 + 2 \theta_m^4 - 5 x^2 \theta_m^2 - 2 x^4 \log(\frac{x}{\theta_m})) \nonumber\\
\chi^{(1)}_{LF}(x)&& /(Y R^2)= \nonumber\\
&&-\frac{1}{64} (x^2 - \theta_m^2)^2 + \frac{s}{16 \pi} (\theta_m^2 - x^2 + 2 x^2 \log(\frac{x}{\theta_m})) \nonumber\\
&&- \frac{1}{2304} (\theta_m^6 + 2 x^6 - 3 x^4 \theta_m^2) \nonumber\\ 
&&+ \frac{s}{384 \pi} (\theta_m^4 - x^2 \theta_m^2 + 2 x^4 \log(\frac{x}{\theta_m})), 
\end{eqnarray}
with $x=r/R$. It is important to note that there are only linear terms in disclination charge $s$, but higher orders in $x$ and $\theta_M$. This is basically due to the fact that defects in both IF and LF appear linearly, but, the displacements do not need to be small. 
The explicit form of the stresses can be found using Eq.~\ref{Eq:form_linear_stress}
\begin{eqnarray}\label{Eq:Results_stress_tar_IF}
&\sigma&^{rr}_{IF}(r)/Y= 
\frac{1}{4} \cos(\frac{r}{R}) 
\left[
-\sec^2(\frac{r}{2R}) + \sec^2(\frac{\theta_m}{2}) \right.\nonumber\\
&&\left.+\frac{s}{2\pi} 
\left( 2\log(\frac{\tan(\frac{r}{2R})}{\tan(\frac{\theta_m}{2})}) 
+ \sec^2(\frac{r}{2R}) - \sec^2(\frac{\theta_m}{2}) \right)
\right], \nonumber \\
&R^2&\sin^2(\frac{r}{R}) \sigma^{\phi\phi}_{IF}(r)/Y=
\frac{1}{4} \cos(\frac{r}{R}) 
\left[
\sec^2(\frac{r}{2R}) + \sec^2(\frac{\theta_m}{2}) \right.\nonumber\\
&&\left. +\frac{s}{2\pi}  
\left(2\log(\frac{\tan(\frac{r}{2R})}{\tan(\frac{\theta_m}{2})}) - \sec^2(\frac{r}{2R}) - \sec^2(\frac{\theta_m}{2})\right)
\right] \nonumber\\ 
&&+ \frac{s}{2\pi} - \frac{1}{2}
\end{eqnarray}
and
\begin{eqnarray}\label{Eq:Results_stress_tar_LF}
&\sigma&^{rr}_{LF}(r)/Y= \nonumber\\
&&\frac{1}{2} \sec^2(\frac{r}{2R}) \cos(\frac{r}{R}) \times \nonumber\\
&&\left[
-\cot^2(\frac{r}{2R}) \log(\cos^2(\frac{r}{2R})) + \cot^2(\frac{\theta_m}{2}) \log(\cos^2(\frac{\theta_m}{2}))\right. \nonumber\\ 
&&\left. + \frac{s}{2\pi} \left(\log(\frac{\tan(\frac{r}{2R})}{\tan(\frac{\theta_m}{2})}) \right.\right. \nonumber\\ 
&&\left.\left.+ \csc^2(\frac{r}{2R}) \log(\cos(\frac{r}{2R})) - \csc^2(\frac{\theta_m}{2}) \log(\cos(\frac{\theta_m}{2}))\right)
\right], \nonumber
\end{eqnarray}
\begin{eqnarray}
&R^2&\sin^2(\frac{r}{R})\sigma^{\phi\phi}_{LF}(r)/Y=1 + \frac{1}{2} \sec^2(\frac{r}{2R}) \times\nonumber\\
&&\left[
\cot^2(\frac{r}{2R}) \log(\cos^2(\frac{r}{2R})) +\cot^2(\frac{\theta_m}{2}) \log(\cos^2(\frac{\theta_m}{2})) \right. \nonumber\\ 
&&\left.+ \frac{s}{2\pi} 
\left(\log(\frac{\tan(\frac{r}{2R})}{\tan(\frac{\theta_m}{2})}) - \cos(\frac{r}{R}) \csc^2(\frac{r}{2R}) \log(\cos(\frac{r}{2R})) \right.\right.\nonumber\\ 
&&\left.\left.- \csc^2(\frac{\theta_m}{2}) \log(\cos(\frac{\theta_m}{2}))\right) \right] \ ,
\end{eqnarray}
which we thoroughly analyze in the next section.

\section{Discussion}\label{Sect_Discussion}

We now present approximate solutions and compare them to those of the exact equations, and analyze each quantity in turn.

\subsection{The function ${\cal F}$}

\begin{figure}
 \centering
  \includegraphics[width=1.0\linewidth]{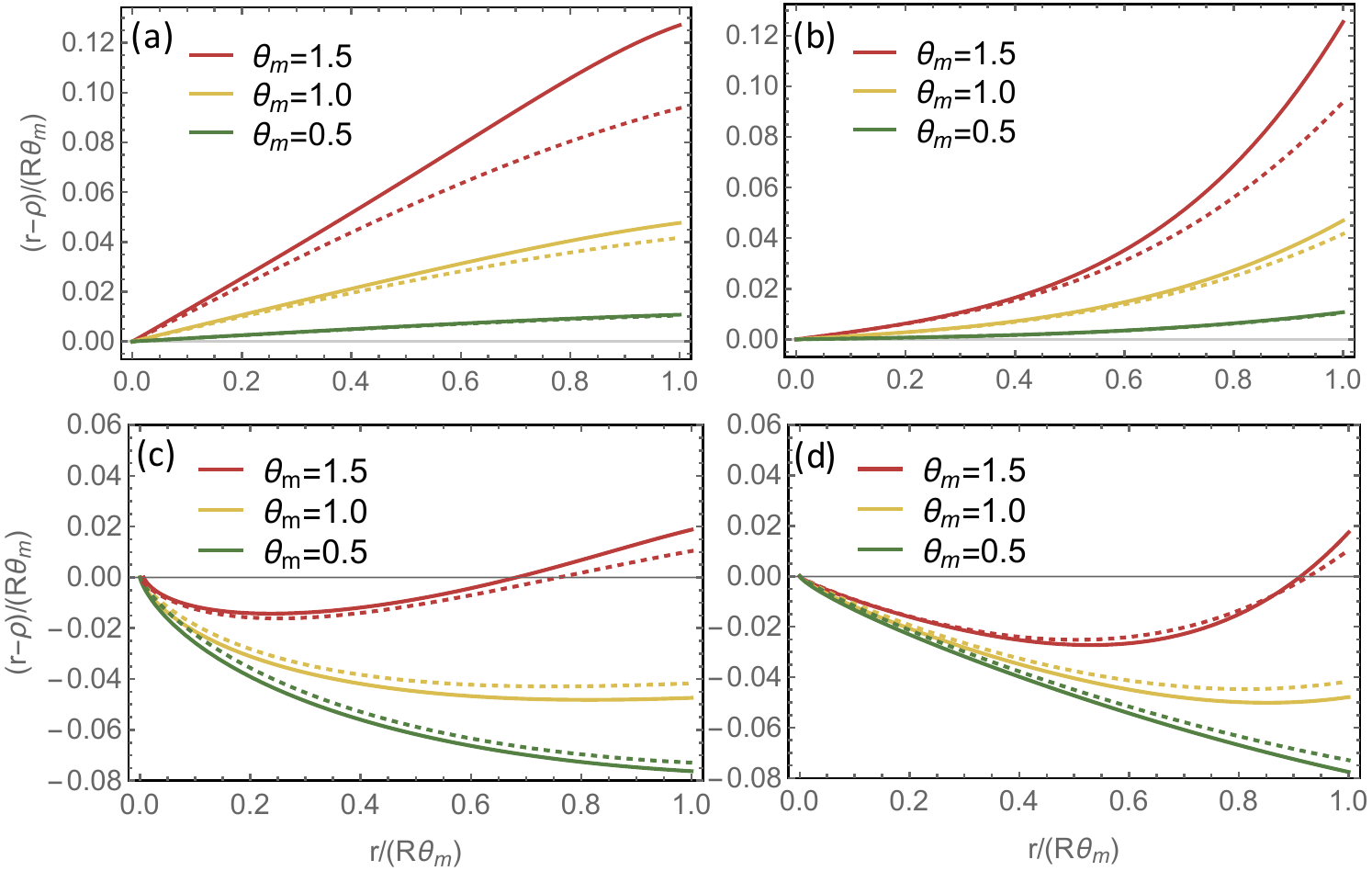}
  \caption{\footnotesize The difference between actual and reference coordinate ($r-\rho(r)$) as a function of the actual coordinate ($r$) for different values of disclination charge $s$ and Poisson ratio $\nu_p$ (a) [$s=0$, $\nu_p=0.2$], (b) [$s=0$, $\nu_p=0.8$], (c) [$s=\frac{\pi}{3}$, $\nu_p=0.2$] and (d) [$s=\frac{\pi}{3}$, $\nu_p=0.8$]. The solid lines correspond to the exact result Eq.~\ref{Eq:Exact_eq_rho} while the dotted lines denote the EF solution Eq.~\ref{Eq:Results_rho_mapping}.}
  \label{compare_map}
\end{figure}

This function defines how distances between particles in reference frame are transformed in actual space. We have not been able to find an analytical expression for the exact Eq.~\ref{Eq:Exact_eq_rho}, which we could nevertheless solve numerically. In Fig.~\ref{compare_map} we compare it to the EF solution defined by Eq.~\ref{Eq:Results_rho_mapping}. In order to visualize the difference, the figures are shown as a function of $r-\rho(r)$. Quite interestingly, the EF mapping shows very small errors, certainly for $\theta_m<0.1$, which corresponds to an aperture angle of 60 degrees. Even for $\theta_m \sim 1.5$ (half the sphere), the linear approximation does extremely well when a disclination is present, which is expected as the disclination charge screens the Gaussian curvature, so that the geometric frustration parameter $\eta$, see Eq.~\ref{Eq:EM_eta}, is small and subsequent corrections to the linear contribution become very small.

\subsection{Airy function and stresses}

\begin{figure}
 \centering
  \includegraphics[width=0.8\linewidth]{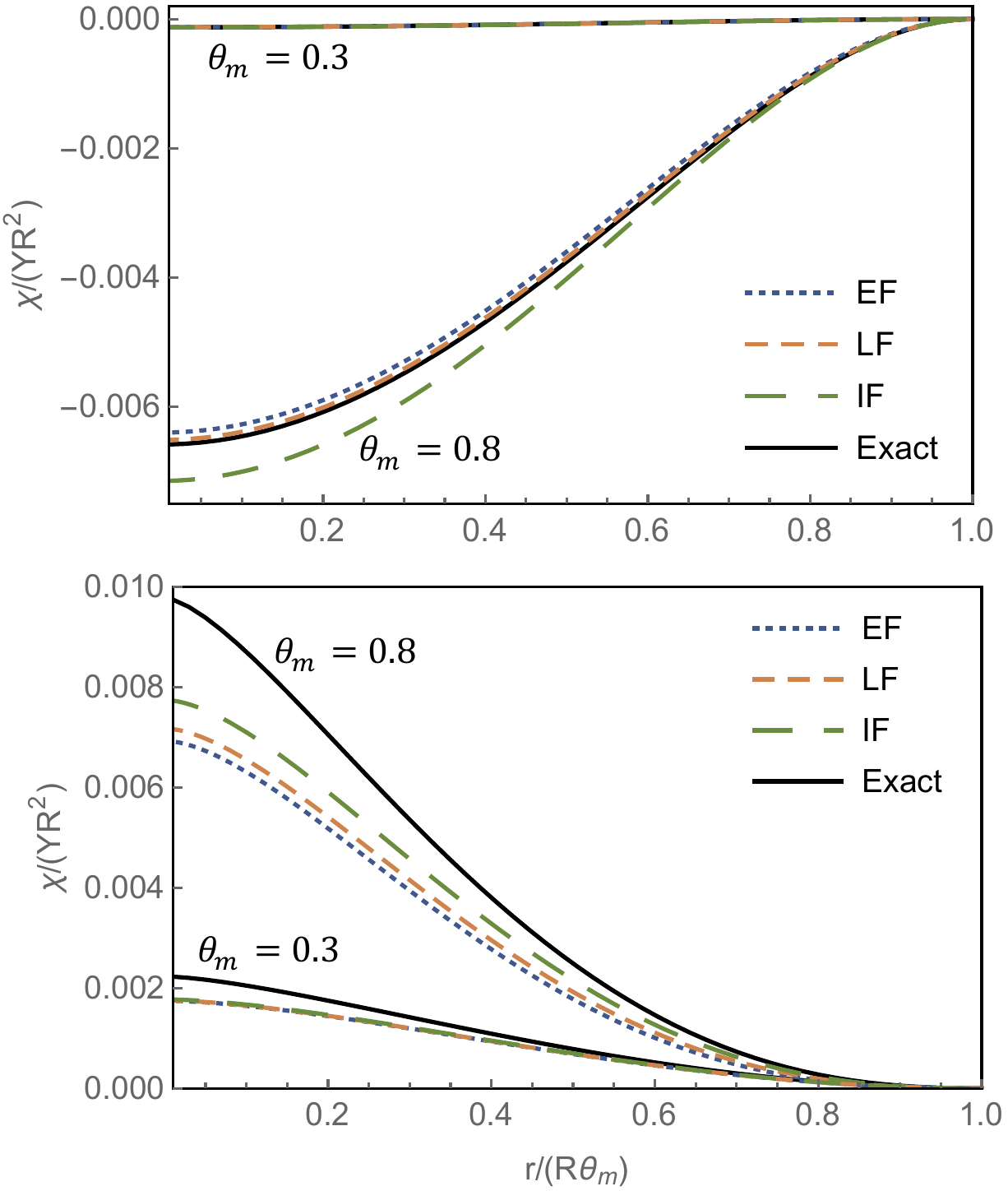}
  \caption{\footnotesize $\chi$ as function of $r$ (actual frame) or $\rho$ (reference frame) corresponding to cap sizes $\theta_m=0.8$ and $\theta_m=0.3$. The upper figure denotes to $s=0$ and lower one with $s=\pi/3$. }
  \label{compare_chi}
\end{figure}

The Airy function, computed with the different approximations, namely EF (Eq.~\ref{Eq:Results_Chi_ref}), IF (Eq.~\ref{Eq:Results_Chi_tar_cur}) and LF (Eq.~\ref{Eq:Results_Chi_tar}) is shown in Fig.~\ref{compare_chi} for two different values of the aperture angle (cap size). Small but significant differences are observed for larger caps.
 
\begin{figure}
 \centering
  \includegraphics[width=1.0\linewidth]{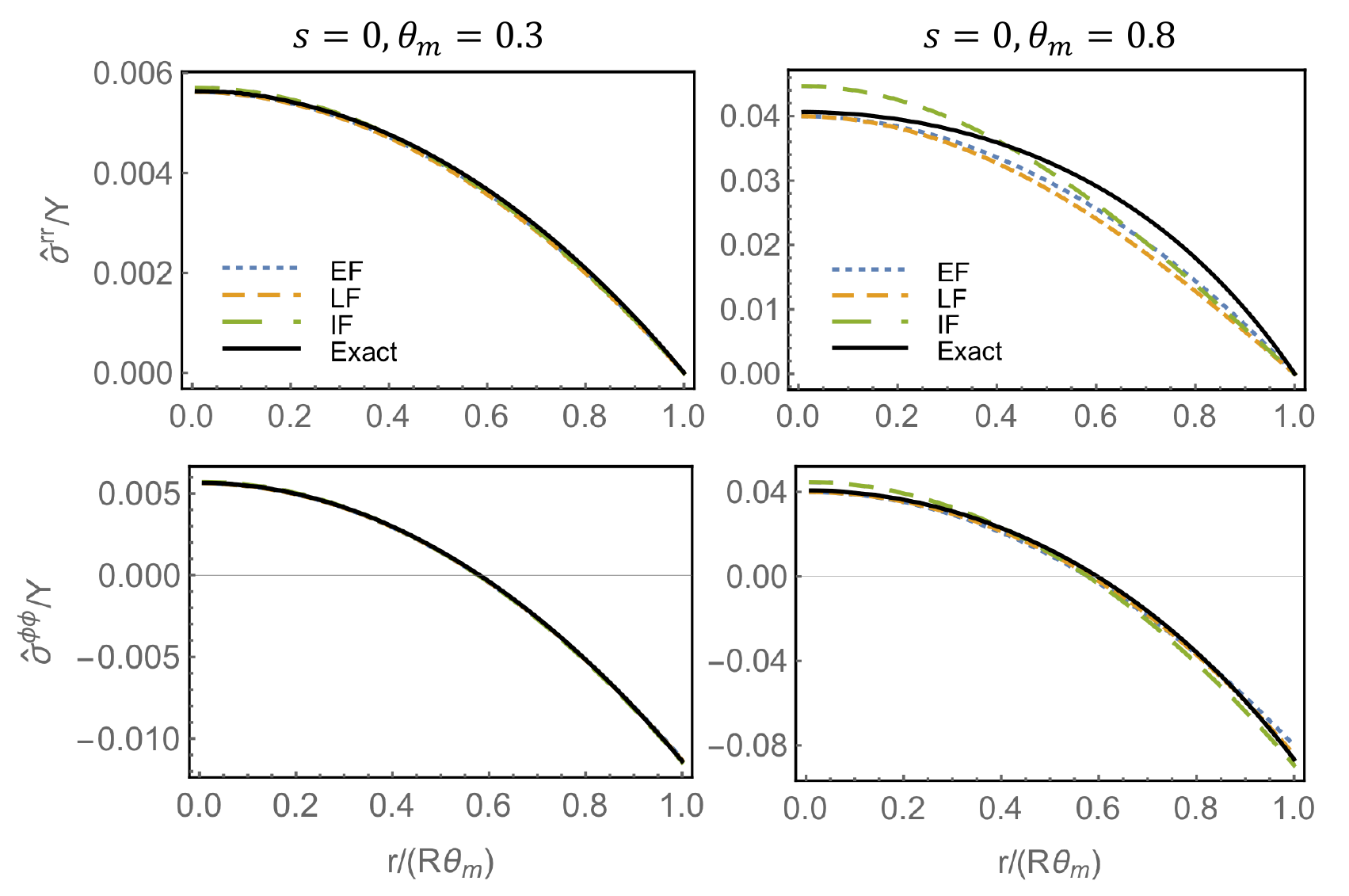}
  \includegraphics[width=1.0\linewidth]{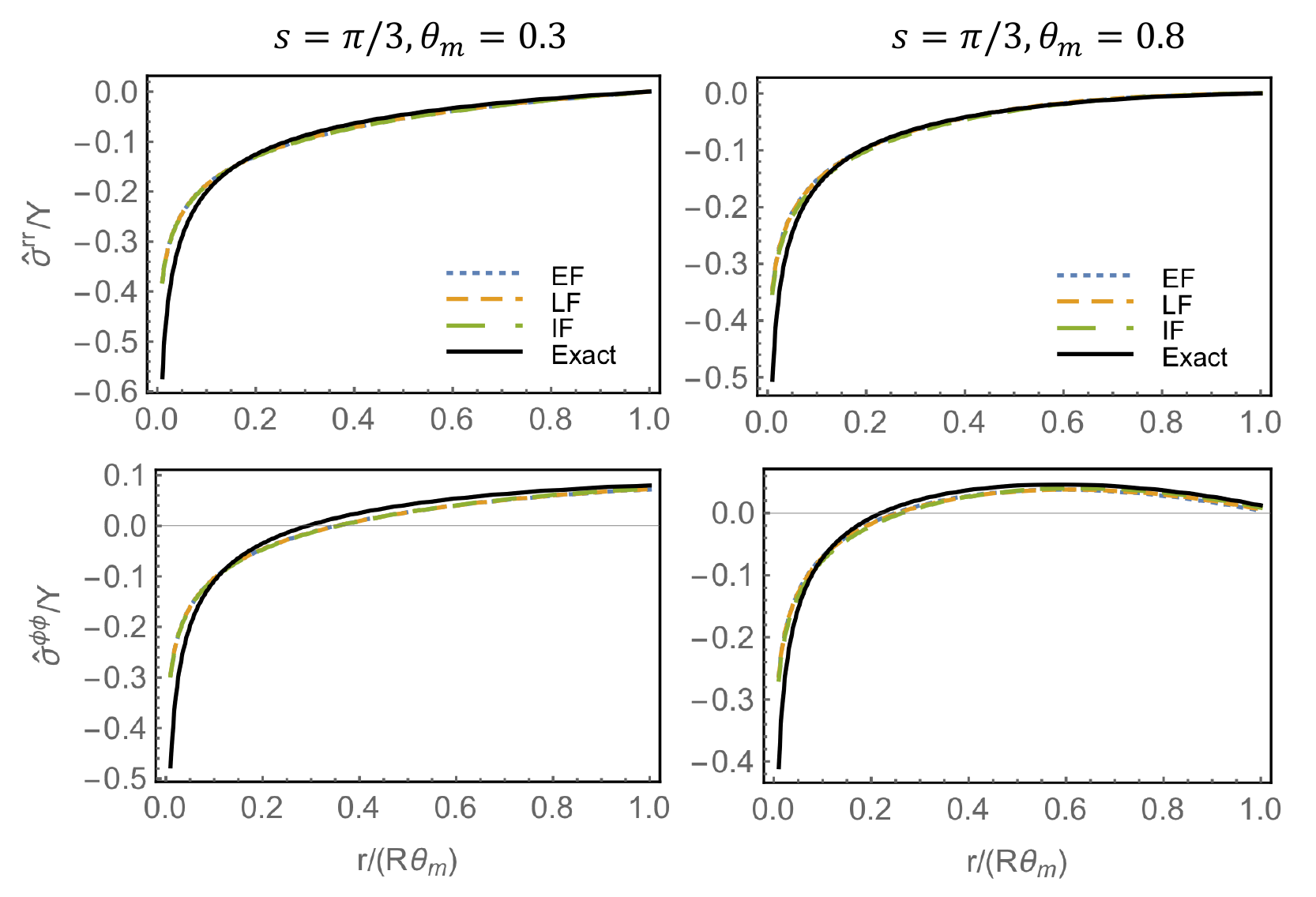}
  \caption{\footnotesize Stress $\sigma^{rr}$ and $\sigma^{\phi\phi}$ with small cap size ($\theta=0.3$, left column) and large cap size ($\theta=1.0$, right column). The top four plots of stress correspond to zero disclination and four bottom plots to a single disclination at the center.}
  \label{stress}
\end{figure}

The stresses show similar trends as observed for the Airy function illustrated in Fig.~\ref{stress}. As expected, for large values of the apeture angle the exact result is in much better agreement with the case of a disclination at the center (note the different scales in the plot).

\subsection{Energy}

\begin{figure}
 \centering
  \includegraphics[width=0.9\linewidth]{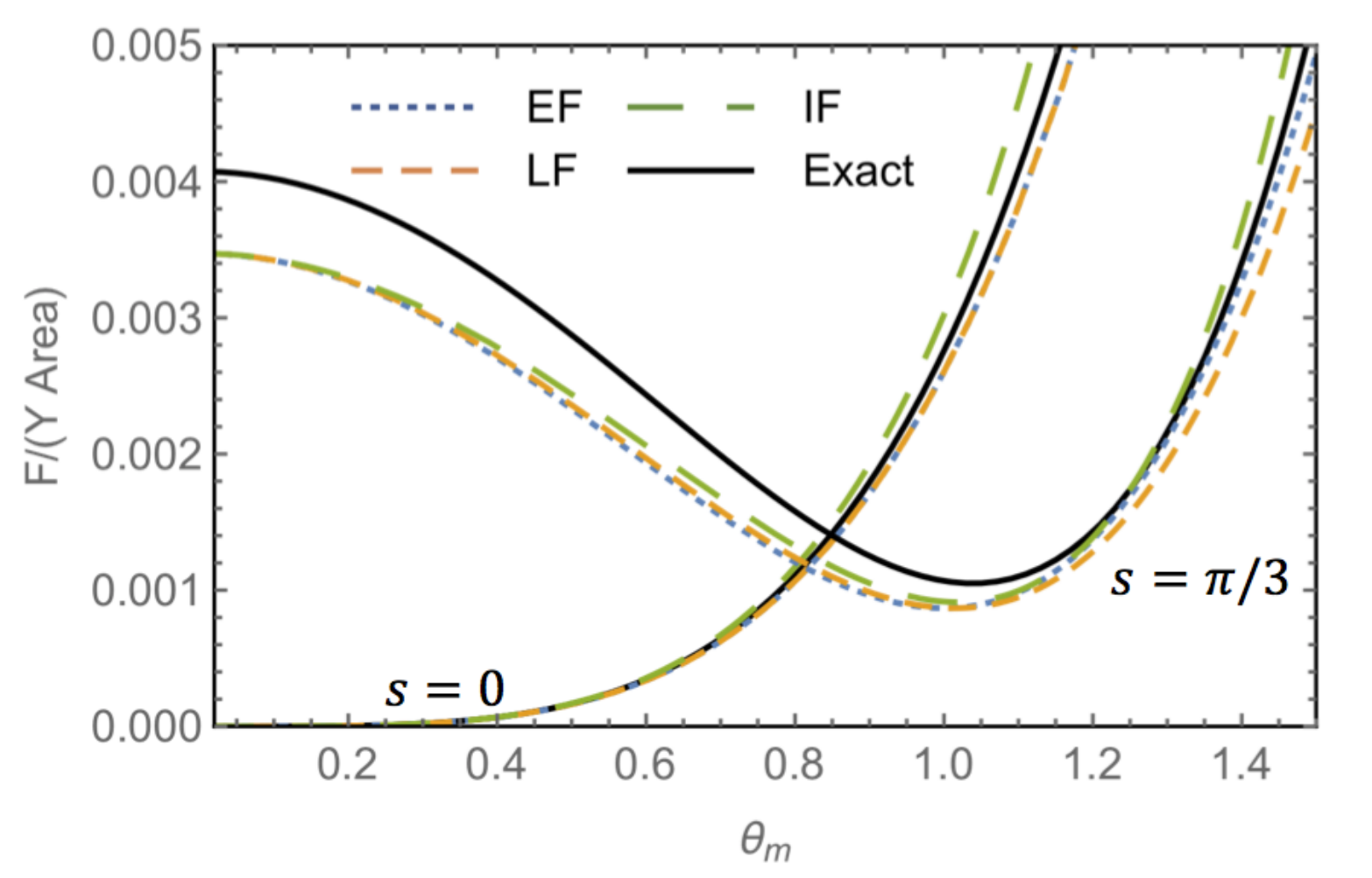}
  \caption{\footnotesize Free energy per unit area for $s=0$ and $s=\frac{\pi}{3}$ for different model presented in the paper.}
  \label{energydensity}
\end{figure}

The values for the total free energy are shown in Fig.~\ref{energydensity} as a function of the aperture angle $\theta_m$ . As expected, in the flat limit $\theta_m \rightarrow 0$, the EF, LF and IF all converge to a value that is different from the exact result, which is also slightly different from another exact result obtained by Seung and Nelson~\cite{Seung1988} (see the discussion in conclusions and appendix), namely
\begin{eqnarray}\label{Eq:Res_NL_Amb}
\frac{F}{Y \mbox{Area}} = \frac{1}{288} &=& 0.0035 \mbox{ (EF, LF, IF)} \\\nonumber
                                &=& 0.0041 \mbox{ (Exact)}   \\\nonumber
                                &=& 0.0040 \mbox{ (Exact SN)} \ .
\end{eqnarray}
The (small) disagreement between EF, LF and IF with the exact result is a consequence of large displacements near the core of a disclination on a flat topography~\cite{TravessetMe2003}. The small disagreement with SN results also reflects the intrinsic ambiguity of what is meant by an ``exact'' elastic theory, as terms with higher powers of the strain tensor, for example, maybe included in the definition of the elastic energy Eq.~\ref{Eq:Form_Elastic}, a point which we will elaborate in the conclusions.

For the case of a central disclination, at finite and increasing values of the aperture angle $\theta_m$, the different linear approximations gradually converge to the exact result. Note that the free energy goes through a minimum at around $\theta_m\approx 1.05$, which maybe interpreted as the point where the disclination optimally screens the Gaussian curvature. It seems reasonable that this point maybe calculated when the PCC Eq.~\ref{Eq:EM_pcc} is satisfied on average, namely 
\begin{equation}
\int d^2{\bf x}s({\bf x})=\int d^2{\bf x}K({\bf x}) \rightarrow \frac{\pi}{3}=2\pi(1-\cos(\theta_c)) \ ,
\end{equation}
that is, at $\theta_M=\theta_c=\arccos(5/6)=0.59$, which is significantly lower and reflects the role of the boundary conditions.  It is also important to note that when $\theta_M > \theta_c$, the approximation to the energy for the disclination free monolayer starts to deviate from the exact result.

\section{Conclusions}\label{Sect_Conclusions}

In this paper we have presented a general fully covariant elastic theory, as defined by the energy Eqs.~\ref{Eq:Form_Elastic} and \ref{Eq:form_W_def}, anticipated in  Refs.~\cite{Efrati2009,MosheSharonKupferman2015}. We discussed three different linear approximations (EF, LF, IF) from which all analytical results quoted in the literature have been derived. Quite unexpectedly, the differences are quantitatively very small, but the ones in actual space (LF, IF) have the advantage that satisfy topological relations, see Eq.~\ref{Eq:EM_Topo}, exactly. It is possible to compute orders beyond linear and, in this way, obtain the exact result, although for general problems, the calculations are quite demanding.

The actual meaning of the ``exact solution'', however, appears as an ambiguous concept.  While our exact result of a single disclination on a flat monolayer as $\theta_m \rightarrow 0$ is almost the same as the value (see Eq.~\ref{Eq:Res_NL_Amb}) obtained by
Seung and Nelson~\cite{Seung1988}, it is not obvious that the energies obtained by the two methods match for all values of $\theta_m$.  The Seung and Nelson's energy is given as
\begin{equation}\label{Eq:Df_elastic_discrete}
F_D = \frac{\epsilon}{2}  \sum_{\langle i, j \rangle} \left(d_{ij}-\bar{d}_{ij}\right)^2
 = \frac{\epsilon}{2} \sum_{\langle i, j \rangle} \left(|{\vec r_i-\vec r_j}|-a\right)^2
\end{equation}
where $\langle i,j \rangle$ are the nearest neighbors defined by a triangulation ${\cal T}$. This energy is conceptually the same as the one defined by Eqs.~\ref{Eq:Form_Elastic} and~\ref{Eq:form_W_def}, since $\bar{d}_{ij}=a$ is the distance in reference and $d_{ij}$ in actual space, and, expanding in small displacements, both energies coincide for the choices of elastic constants $Y=2\epsilon/\sqrt{3}$ and $\nu_p=1/3$~\cite{Seung1988}. However, these two approaches differ beyond linear order. It is possible to make them agree at higher orders by adding  higher powers of $|d_{ij}-\bar{d}_{ij}|$ in Eq.~\ref{Eq:Df_elastic_discrete} \ ,
\begin{equation}\label{Eq:Df_elastic_discrete_more}
F = F_D + \sum_{l=2}^M \frac{\epsilon_l}{2}\left(d_{ij}-\bar{d}_{ij}\right)^{2l}
\end{equation}
so that, for appropriately chosen values $\epsilon_l$, higher orders of the displacement beyond linear will agree with the energy Eq.~\ref{Eq:form_W_def}. Additional powers of $u_{\alpha \beta}$ can also be added to Eq.~\ref{Eq:form_W_def}, to make it agree with  Eq.~\ref{Eq:Df_elastic_discrete}. Either case, it serves to make the point that Eqs.~\ref{Eq:form_W_def} and \ref{Eq:Df_elastic_discrete} represent two different non-linear elastic theories, and therefore, it is expected that the exact results for a single disclination will differ. It should be noted, however, that both exact results are close, thus highlighting that non-linear corrections are small. The natural question becomes then, which one is the ``correct'' model. A satisfactory answer can be given if the underlying microscopic potential among particles is known. Then it is possible to impose that the higher orders of elasticity theory (see Eq.~\ref{Eq:Df_elastic_discrete_more}) match the same orders of the energy of the crystal in powers of the displacement, as discussed in Ref.~\cite{BowickMea2006}, where exceedingly accurate predictions for energies were obtained for any geometry. 

\begin{figure}
	\centering
	\includegraphics[width=1.0\linewidth]{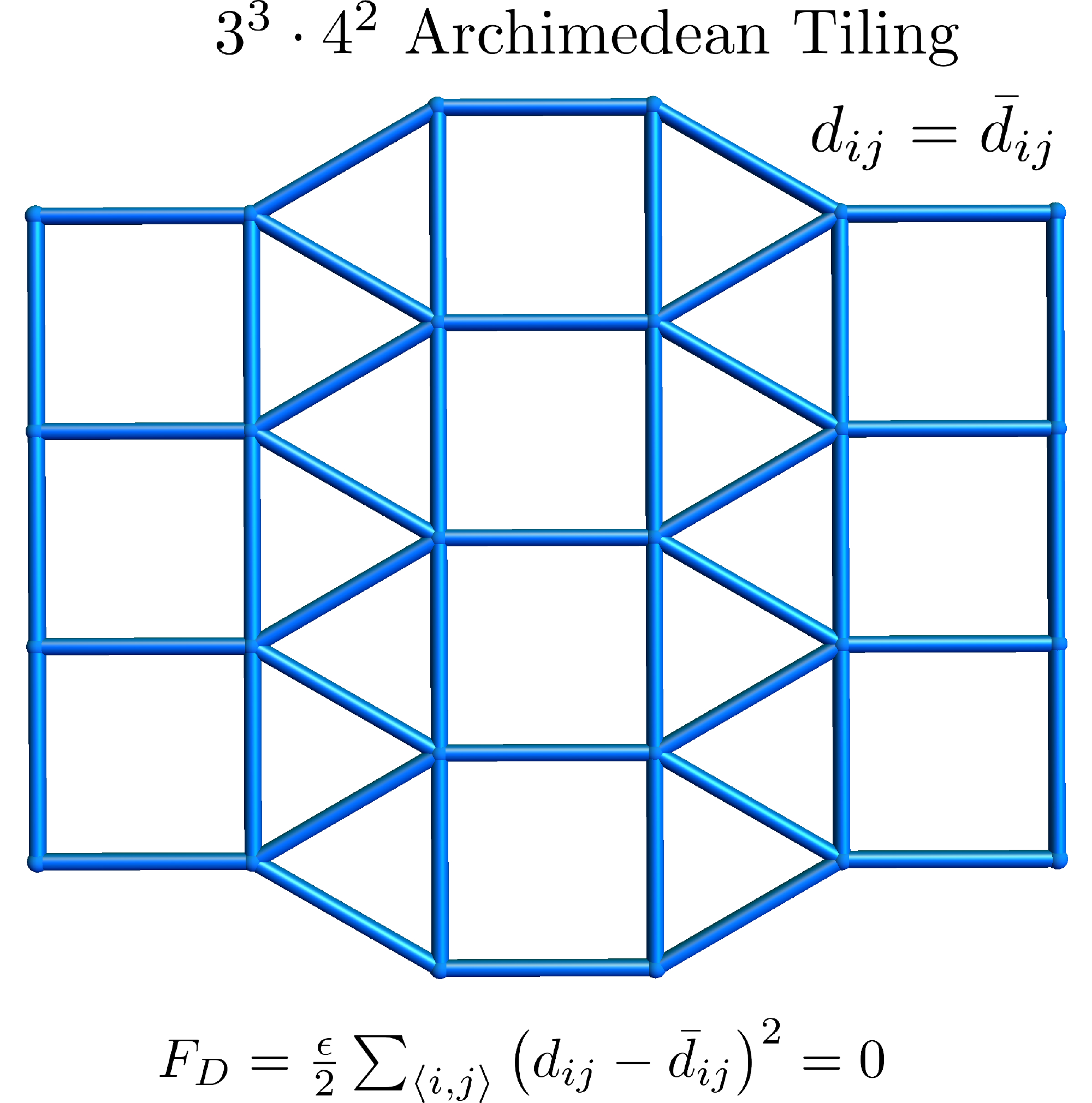}
	\caption{Example of the $(3^3.4^2)$ Archimedean tiling with zero elastic energy. Such configuration, however, has zero energy modes and require additional constraints to be stable.}
	\label{Fig:Arch}
\end{figure}

Another fundamental aspect of the geometric theory of elasticity discussed in this paper is the choice of the reference metric, which corresponds to a configuration where all nearest neighbors distances and angles are the same. In some cases, such as for a defect free disk or a cone with a single disclination, it is possible to optimize the geometry resulting into strain and stress free configurations in actual space. For other, more complex defect distributions, such actual space configurations do not exist. A conspicuous property of the model in Eq.~\ref{Eq:Df_elastic_discrete}, however, is that it involves nearest neighbor distances only, and the condition that the angles are the same does not need to be satisfied. Thus, general Archimedean tilings configurations, such as the one shown in Fig.~\ref{Fig:Arch}, are strain/stress free for a actual space consisting of a plane. It is interesting to note that it is possible to build dodecagonal quasicrystals out of $(3^3.4^2)$ Archimedean tiling, which have been observed in nanocrystal systems~\cite{Talapin2009a}. Within elasticity theory, those Archimedean tilings require a Poisson ratio $\nu_p=1/3$, as clear from the discussion following Eq.~\ref{Eq:Df_elastic_discrete}, see also Ref.~\cite{TravessetMe2003}. 

We have shown that the ``exact'' equations of elasticity theory amount to minimizing the difference between the actual and the reference metric
\begin{eqnarray}\label{Eq:Concl_strain}
g(\mbox{actual metric}) - \bar{g}(\mbox{Reference metric}) &=& 2 u_{\alpha \beta} \nonumber
\end{eqnarray}
where the actual metric is fixed by the topography (the surface), see Fig.~\ref{Fig:Ref_vsTarg}, and the reference metric is such that its curvature $\bar{K}$ is a sum of disclinations and dislocations
\begin{eqnarray}\label{Eq:Concl_metric} 
\bar{K} &=& \mbox{Disclinations} + \mbox{Dislocations} \nonumber \\
&=& \mbox{``Quanta'' of Curvature} + \mbox{``Quanta'' of Torsion} \ ,
\nonumber
\end{eqnarray}
where the disclinations are quantized in units of $\frac{\pi}{3}$ and the dislocations in units of the Burgers vector ${\vec b}$. These equations summarize the geometric content of the equations in elasticity theory as applied to arbitrary topographies. For boundary free crystals, they also satisfy topological constraints, for example, Eq.~\ref{Eq:EM_Topo}.

There are a number of issues that we have not discussed. For example, the free energy Eq.~\ref{Eq:form_W_def} is invariant under general parameterizations, which in turns, through the Noether theorem, gives rise to conservation laws that relate to the stress tensor. Also, the IF includes a term, see Eq.~\ref{Eq:form_linear_scalar}, that has derivative of the Gaussian curvature. In those cases where the Gaussian curvature is not constant and varies rapidly, this term may become important or even dominant.

In summary, we presented a covariant formulation of elasticity that unifies geometric and topological concepts with the theory of defects. All available results in the literature maybe recovered from this formulation as suitable approximations, thus providing a rigorous justification on their validity, and providing the necessary framework for our recent studies of icosahedral order in virus shells~\cite{LiMe2018}. Throughout this paper, the geometry has been fixed. There are obviously many fascinating problems when the geometry is allowed to fluctuate, see, for example Ref.~\cite{Callens2018}, but those problems will be discussed elsewhere.

\begin{acknowledgments}
	We are grateful to Greg Grason for many discussions and insightful comments as well as for his carefully reading the mansucript.   The SL and RZ were supported by NSF Grant No. PHY-1607749 and AT was supported by NSF, DMR-CMMT 1606336.
\end{acknowledgments}

\appendix

\section{The Seung-Nelson result as a function of area}

Seung-Nelson~\cite{Seung1988} quote, for a flat disclination
\begin{equation}
\frac{F}{Y s^2 R^2}= 0.008 \ .
\end{equation}
The radius is given by $R= n a$, where $n$ is an integer and $a$ is the lattice constant. A more precise calculation computes this coefficient as $0.00785$~\cite{TravessetMe2003}. This is a numerical calculation considering a pentagonal shape crystal containing $5 n^2$ triangles. Each trianlge has an area $\frac{\sqrt{3}}{4} a^2$, hence 
\begin{equation}
\frac{F}{Y Area}= 0.008\left(\frac{\pi}{3}\right)^2 /(5\sqrt{3}/4)\approx 0.00405  \ ,
\end{equation}
or $0.00400$ with the more precise value~\cite{TravessetMe2003}. This is the coefficient used in Eq.~\ref{Eq:Res_NL_Amb}.

\section{Geometry, curvature, vielbeins and the definition of the stress tensor}\label{App_Vielbeins}

It should be noticed that the stress tensor, defined by Eq.~\ref{Eq:form_Stress_tensor} is in general different than the one defined in standard textbooks, such as Landau and Lifshitz, which we denote
as $\hat{\sigma}^{\alpha \beta}$. We now show the relation between both tensors. For that purpose, we introduce the Vielbeins $e\indices{^\alpha_\mu}$, defined as
\begin{eqnarray}\label{Eq:App_Veilbeins}
g_{\mu \nu} &=& e\indices{^\alpha_\mu}e\indices{^\beta_\mu}\delta_{\alpha\beta} \nonumber\\
\delta_{\alpha \beta} &=& e\indices{_\alpha^\mu}e\indices{_\beta^\mu}g_{\mu\nu}
\end{eqnarray} 
Then, there is the relation
\begin{equation}\label{Eq:App_stress_stress}
\hat{\sigma}^{\alpha \beta} = e\indices{^\alpha_\mu}e\indices{^\beta_\nu}\sigma^{\mu \nu} \ .
\end{equation}
The advantage of $\hat{\sigma}^{\alpha \beta}$ is that the units of all the components are the same. This is not the case for $\sigma^{\mu \nu}$. Obvious to say that all physical quantities have the same dimensions in either form.

Also, the line tension term Eq.~\ref{Eq:form_line_tension} is simplified by
\begin{equation}\label{Eq:App_line_tension}
\int_{\partial\cal B} ds = \int_{\partial\cal B} \sqrt{g} dl = \int_{\partial\cal B} dx^{\mu} g_{\mu\nu} t^{\nu}\ ,
\end{equation}
where $t^{\nu}=\frac{1}{\sqrt{g}} \frac{d x^{\mu}}{d l}$ for any parameterization $x^{\mu}(l)$. Here $t^{\mu}$ is the unit tangent vector to the curve defining the boundary. Note that 
\begin{equation}\label{Eq:App_line_tension_new}
g = g_{\mu \nu} \frac{d x^{\mu}}{dl} \frac{d x^{\nu}}{dl}
\end{equation}
and $dx^{\mu}= \sqrt{g} t^{\mu} dl$. The variation of this term gives
\begin{eqnarray}\label{Eq:App_line_der_tension}
\int_{\partial\cal B} dx^{\mu} \delta g_{\mu\nu} t^{\nu} &=&
-\int_{\partial\cal B} dx^{\mu} (\nabla_{\nu} \xi_{\mu}+ \nabla_{\mu} \xi_{\nu})t^{\nu} \nonumber \\
&=& \int_{\partial\cal B} dx^{\mu} (\xi_{\mu} \nabla_{\nu} + \xi_{\nu}\nabla_{\mu})t^{\nu} \nonumber \\
&=& \int_{\partial\cal B} dx^{\mu} \nabla_{\mu} t^{\nu} \xi_{\nu} \ ,
\end{eqnarray}
where $dx^{\mu}\xi_{\mu}=0$ as the vector $\xi^{\mu}$ is perpendicular to $t^{\mu}$. Note that the vector
\begin{equation}\label{Eq:App_line_perp}
n_{\rho}=\sqrt{g} \epsilon_{\mu \rho} t^{\mu} \ ,
\end{equation}
is a unit vector, perpendicular to $t^{\mu}$.

The variation in Eq.~\ref{Eq:App_line_der_tension} refers to 
$\delta g_{\alpha \beta}$ with the implicit condition $\delta {\bar g}_{\alpha \beta}=0$, while the variation leading to Eq.~\ref{Eq:form_variation} is with respect to $\delta {\bar g}_{\alpha \beta}$ with $\delta g_{\alpha \beta}=0$. One notes, however, that the general transformation
\begin{eqnarray}
\delta g_{\alpha \beta} &=& \nabla_{\alpha} \xi_{\beta}+ \nabla_{\beta} \xi_{\alpha} \nonumber \\ 
 \delta {\bar g}_{\alpha \beta} &=& {\bar \nabla}_{\alpha} \xi_{\beta}+ {\bar \nabla_{\beta}} \xi_{\alpha} \ ,
\end{eqnarray}
encodes a simple reparamaterization and therefore, under this transformation any term $F_a$ appearing in the energy should satisfy
\begin{equation}
    \delta F_a =  \delta_{g} F_a + \delta_{\bar g}F_a=0\ ,
\end{equation}
hence, the correct variation, with respect to $\bar{g}_{\alpha \beta}$ picks up a minus sign, as compared with Eq.~\ref{Eq:App_line_der_tension}, 
\begin{equation}
 \delta F_l =-\int_{\partial\cal B} dx^{\mu} \nabla_{\mu} t^{\nu} \xi_{\nu} \ ,
\end{equation}
as used in the main text.

\section{Incompatibility metric approximations}\label{App_IM}

\subsection{Incompatibility metric approximation: actual frame}

The second order in the expansion Eq.~\ref{Eq:form_expansion} is given by
\begin{eqnarray}\label{Eq:form_expansion_second}
g\indices*{_\alpha_\beta^{(2)}} &=& -\frac{2}{Y}\left( g_{\alpha \beta} \Delta \chi^{(2)} - (1+\nu_P)\nabla_{\alpha}\nabla_{\beta} \chi^{(2)} \right) \nonumber\\
&-& \frac{2}{Y}\left(g_{\alpha \beta} g^{\rho \gamma} \tensor*[]{\Gamma}{^{\mu}_\rho_\gamma^{(1)}}  - (1+\nu_P)\tensor*[]{\Gamma}{^{\mu}_\alpha_\beta^{(1)}}  \right)\nabla_{\mu} \chi^{(1)} \nonumber\\
&-&\frac{1}{2}  g\indices*{_\alpha_\beta^{(1)}} g^{\gamma \sigma} g\indices*{_\gamma_\sigma^{(1)}}
\end{eqnarray}
Obviously, the expansion can be continued to all orders, and in this way a perturbative solution to Eq.~\ref{Eq:form_elastic_equation_m} and Eq.~\ref{Eq:form_elastic_equation_m_reference} can be found. The goal is now to derive an explicit equation for $\chi^{(i)}$, as shown below.

\subsection{Incompatibility metric approximation: reference frame}

The second order in Eq.~\ref{Eq:form_orders_ref} can also be computed as:
\begin{eqnarray}\label{Eq:form_expansion_second_ref}
{\bar g}\indices*{_\alpha_\beta^{(II)}} &=& \frac{2}{Y}\left( {\bar g}_{\alpha \beta} {\bar \Delta} \chi^{(II)} - (1+\nu_P){\bar \nabla}_{\alpha}{\bar \nabla}_{\beta} \chi^{(II)} \right) \nonumber\\
&+&\frac{1}{2}  {\bar g}\indices*{_\alpha_\beta^{(I)}} {\bar g}^{\gamma \sigma} {\bar g}\indices*{_\gamma_\sigma^{(I)}}
\end{eqnarray}

\subsection{First order solution: actual frame}

We will compute the Ricci tensor ${\bar R}_{\sigma\nu} = {\bar R}\indices*{^\rho_{\sigma}_\rho_{\nu}}$, which from Eq.~\ref{Eq:App_Riemann_expansion} is
\begin{equation}\label{Eq:form_linear_ricci}
{\bar R}_{\sigma\nu} = R_{\sigma\nu} + \nabla_{\mu} \tensor*[]{\Gamma}{^{\mu}_\nu_\sigma^{(1)}} - \nabla_{\nu} \tensor*[]{\Gamma}{^{\mu}_\mu_\sigma^{(1)}} \ .
\end{equation}
The first term is obtained from Eq.~\ref{Eq:form_orders}, Eq.~\ref{Eq:App_second} and Eq.~\ref{Eq:App_Riemann_expansion}, leading to
\begin{eqnarray}\label{Eq:form_linear_ft}
-Y \nabla_{\mu} \tensor*[]{\Gamma}{^{\mu}_\nu_\sigma^{(1)}} &=&  \nabla_{\sigma}\nabla_{\nu} \Delta \chi^{(1)} -(1+\nu_P) g^{\mu\gamma} \nabla_{\mu} \nabla_{\nu}\nabla_{\sigma}\nabla_{\gamma} \chi^{(1)} \nonumber \\
&+&  \nabla_{\nu}\nabla_{\sigma} \Delta \chi^{(1)} -(1+\nu_P) g^{\mu\gamma} \nabla_{\mu} \nabla_{\sigma}\nabla_{\nu}\nabla_{\gamma} \chi^{(1)}
\nonumber \\
&-&g_{\sigma\nu} \Delta^2 \chi^{(1)} + (1+\nu_P) g^{\mu\gamma} \nabla_{\mu} \nabla_{\gamma}\nabla_{\nu}\nabla_{\sigma} \chi^{(1)} \nonumber \\
&&
\end{eqnarray}
This is simplified by using Eq.~\ref{Eq:App_Riemann_2} and Eq.~\ref{Eq:App_Riemann_3}
\begin{eqnarray}\label{Eq:form_linear_last_1}
g^{\mu\gamma} \nabla_{\mu} \nabla_{\gamma}\nabla_{\nu}\nabla_{\sigma} \chi^{(1)}  &=& g^{\mu\gamma} \nabla_{\mu} \nabla_{\nu}\nabla_{\gamma}\nabla_{\sigma} \chi^{(1)} \nonumber \\
&-& g^{\mu\gamma} \nabla_{\mu}\left(R\indices{^\lambda_\sigma_\gamma_\nu} \nabla_{\lambda}\chi^{(1)}\right)
\end{eqnarray}
and
\begin{eqnarray}\label{Eq:form_linear_last_2}
g^{\mu\gamma} \nabla_{\mu} \nabla_{\sigma}\nabla_{\nu}\nabla_{\gamma}  \chi^{(1)}  &=& g^{\mu\gamma} \nabla_{\sigma} \nabla_{\mu}\nabla_{\nu}\nabla_{\gamma} \chi^{(1)} \nonumber \\
&-& g^{\mu\gamma} R\indices{^\lambda_\nu_\mu_\sigma}\nabla_{\lambda}\nabla_{\gamma}\chi^{(1)} \nonumber\\
&-& g^{\mu\gamma} R\indices{^\lambda_\gamma_\mu_\sigma}\nabla_{\nu}\nabla_{\lambda}\chi^{(1)} \ .
\end{eqnarray}
One more application of Eq.~\ref{Eq:App_Riemann_3} converts Eq.~\ref{Eq:form_linear_last_2} into
\begin{eqnarray}\label{Eq:form_linear_last_3}
g^{\mu\gamma} \nabla_{\mu} \nabla_{\sigma}\nabla_{\nu}\nabla_{\gamma}  \chi^{(1)}  &=& g^{\mu\gamma} \nabla_{\sigma} \nabla_{\nu}\nabla_{\mu}\nabla_{\gamma} \chi^{(1)} \nonumber \\
&-& g^{\mu\gamma} \nabla_{\sigma}\left(R\indices{^\lambda_\gamma_\mu_\nu} \nabla_{\lambda}\chi^{(1)}\right) \nonumber\\
&-& g^{\mu\gamma} R\indices{^\lambda_\nu_\mu_\sigma}\nabla_{\lambda}\nabla_{\gamma}\chi^{(1)} \nonumber\\
&-& g^{\mu\gamma} R\indices{^\lambda_\gamma_\mu_\sigma}\nabla_{\nu}\nabla_{\lambda}\chi^{(1)} \ .
\end{eqnarray}
Using the expression of the Riemann tensor in two dimensions Eq.~\ref{Eq:App_Riemann_7}, we obtain
\begin{eqnarray}\label{Eq:form_linear_riemann1}
g^{\mu\gamma} R\indices{^\lambda_\nu_\mu_\sigma}\nabla_{\lambda}\nabla_{\gamma}\chi^{(1)} &=& K g_{\nu\sigma} \Delta \chi^{(1)}-K\nabla_{\sigma}\nabla_{\nu} \chi^{(1)} \nonumber\\
g^{\mu\gamma} R\indices{^\lambda_\gamma_\mu_\sigma}\nabla_{\nu}\nabla_{\lambda}\chi^{(1)} &=& - K \nabla_{\nu}\nabla_{\sigma}\chi^{(1)} \ .
\end{eqnarray}
and
\begin{eqnarray}\label{Eq:form_linear_riemann2}
g^{\mu\gamma} \nabla_{\sigma}\left(R\indices{^\lambda_\gamma_\mu_\nu} \nabla_{\lambda}\chi^{(1)}\right) &=& -\nabla_{\sigma}K \nabla_{\nu}\chi^{(1)}-K\nabla_{\sigma}\nabla_{\nu} \chi^{(1)} \nonumber \\
&&\ .
\end{eqnarray}
Also
\begin{eqnarray}\label{Eq:form_linear_riemann3}
g^{\mu\gamma} \nabla_{\mu}\left(R\indices{^\lambda_\sigma_\gamma_\nu} \nabla_{\lambda}\chi^{(1)}\right) &=& g_{\sigma\nu} g^{\mu\lambda}\nabla_{\mu}K \nabla_{\lambda} \chi^{(1)} \nonumber\\
&-& \nabla_{\sigma}K \nabla_{\nu} \chi^{(1)} \nonumber\\
&+& g_{\sigma\nu} K \Delta \chi^{(1)} \nonumber\\
&-& K \nabla_{\sigma}\nabla_{\nu}\chi^{(1)}
\end{eqnarray}
Collecting all these terms, Eq.~\ref{Eq:form_linear_ft}
\begin{eqnarray}\label{Eq:form_linear_ft_final}
-Y \nabla_{\mu} \tensor*[]{\Gamma}{^{\mu}_\nu_\sigma^{(1)}} &=&  2 \nabla_{\sigma}\nabla_{\nu} \chi^{(1)}-g_{\sigma\nu} \Delta^2 \chi^{(1)} -\nonumber\\
&-&(1+\nu_P)\left[\nabla_{\sigma}\nabla_{\nu} \Delta \chi^{(1)}+ 2K\nabla_{\sigma}\nabla_{\nu}\chi^{(1)} + \right.\nonumber\\
&+& \left. g_{\sigma \nu} g^{\mu\lambda} \nabla_{\mu}K \nabla_{\lambda} \chi^{(1)} \right]
\end{eqnarray}
The next quantity to compute is
\begin{eqnarray}\label{Eq:form_linear_fu}
-Y \nabla_{\nu} \tensor*[]{\Gamma}{^{\mu}_\mu_\sigma^{(1)}} &=&  \nabla_{\nu}\nabla_{\sigma} \Delta \chi^{(1)} -(1+\nu_P) g^{\mu\gamma} \nabla_{\nu} \nabla_{\mu}\nabla_{\sigma}\nabla_{\gamma} \chi^{(1)} \nonumber \\
&+& 2 \nabla_{\nu}\nabla_{\sigma} \Delta \chi^{(1)} -(1+\nu_P) g^{\mu\gamma} \nabla_{\nu} \nabla_{\sigma}\nabla_{\mu}\nabla_{\gamma} \chi^{(1)}
\nonumber \\
&-&\nabla_{\nu}\nabla_{\sigma} \Delta\chi^{(1)} + (1+\nu_P) g^{\mu\gamma} \nabla_{\nu} \nabla_{\mu}\nabla_{\gamma}\nabla_{\sigma} \chi^{(1)} \nonumber \\
&&
\end{eqnarray}
that immediately leads to
\begin{eqnarray}\label{Eq:form_linear_fu_final}
-Y \nabla_{\nu} \tensor*[]{\Gamma}{^{\mu}_\mu_\sigma^{(1)}} &=& 2\nabla_{\sigma}\nabla_{\nu}\Delta \chi^{(1)} - (1+\nu_P)\nabla_{\nu}\nabla_{\sigma} \chi^{(1)}
\end{eqnarray}
Therefore, the Ricci tensor is
\begin{eqnarray}\label{Eq:form_lienar_ricci_ex}
{\bar R}_{\sigma \nu} &=& R_{\sigma \nu}+\frac{1}{Y}\left( g_{\sigma\nu} \Delta^2 \chi^{(1)} + \right.\nonumber\\
&+&\left.(1+\nu_P)\left[2K\nabla_{\sigma}\nabla_{\nu}\chi^{(1)} + g_{\sigma \nu} g^{\mu\lambda} \nabla_{\mu}\chi^{(1)}\nabla_{\lambda} K \right]\right) \nonumber\\
&&
\end{eqnarray}
Finally, the scalar curvature is obtained as the trace of the Ricci tensor, hence
\begin{eqnarray}\label{Eq:form_linear_scalar_app}
{\bar K} &=& K +\frac{1}{Y} \left(\Delta^2 \chi^{(1)} + \right. \\\nonumber
&+& \left. 2K \Delta\chi^{(1)}+(1+\nu_p) g^{\mu\lambda} \nabla_{\mu}K \nabla_{\lambda} \chi^{(1)}\right) \ .
\end{eqnarray}

\section{Elastic energy in the actual frame}\label{App_Energy_Target}

Our starting point is Eq.~\ref{Eq:form_linear_free_energy}, which for the sake of reference we repeat here:
\begin{eqnarray}\label{Eq:App_tf_form_linear_free_energy}
F & = & \frac{1}{2Y} \int d^2u \sqrt{g} \left[ (\Delta \chi^{(1)})^2 +\right. \nonumber \\
&+&\left. \frac{(1+\nu_P)}{g} \epsilon^{\alpha \sigma}\epsilon^{\rho \beta} \nabla_{\alpha}\nabla_{\beta} \chi^{(1)} \nabla_{\rho}\nabla_{\sigma} \chi^{(1)}\right] .
\end{eqnarray}
We now focus on the second term. Using Eq.~\ref{Eq:App_Riemann_5} this term becomes
\begin{eqnarray}\label{Eq:App_tf_id1}
& &\epsilon^{\alpha \sigma}\epsilon^{\rho \beta} \nabla_{\alpha}\nabla_{\beta} \chi^{(1)} \nabla_{\rho}\nabla_{\sigma} \chi^{(1)} \\ \nonumber
&=& g\left[ \nabla_{\alpha}\nabla_{\beta} \chi^{(1)} \nabla^{\alpha}\nabla^{\beta} \chi^{(1)} -(\Delta \chi^{(1)})^2\right] \ .
\end{eqnarray}
Making  further use of Eq.~\ref{Eq:App_Riemann_0}, allows to prove the following identity
\begin{eqnarray}
& & \sqrt{g} T^{\alpha \beta}\nabla_{\alpha}\nabla_{\beta} \chi^{(1)} \\\nonumber
&=& \partial_{\alpha}\left(\sqrt{g}T^{\alpha \beta}\nabla_{\beta}\chi^{(1)}\right)-
\sqrt{g}\nabla_{\alpha}T^{\alpha\beta}\nabla_{\beta} \chi^{(1)}\\  \nonumber
&=& \partial_{\alpha}\left(\sqrt{g}T^{\alpha \beta}\nabla_{\beta}\chi^{(1)}\right)-\sqrt{g}\nabla_{\alpha}(g^{\alpha\beta}\Delta \chi^{(1)})\nabla_{\beta}\chi^{(1)} -\\\nonumber
&-& \sqrt{g} K  g^{\beta \alpha} \nabla_{\alpha} \chi^{(1)} \nabla_{\beta} \chi^{(1)} 
\end{eqnarray}
where $T^{\alpha \beta}=\nabla^{\alpha}\nabla^{\beta} \chi^{(1)} $. Note that
\begin{eqnarray}
\nabla_{\alpha} T^{\alpha \beta} &=& \nabla_{\alpha}g^{\alpha\rho}g^{\beta \nu}\nabla_{\rho}\nabla_{\nu} \chi^{(1)}= g^{\beta \nu}g^{\alpha\rho} \nabla_{\alpha}\nabla_{\rho}\nabla_{\nu} \chi^{(1)} \nonumber\\
&=& g^{\beta \nu} g^{\alpha\rho} \nabla_{\nu} \nabla_{\alpha}\nabla_{\rho} \chi^{(1)} - g^{\beta \nu} g^{\alpha\rho}  R\indices{^\lambda_{\rho}_\alpha_\nu}\nabla_{\lambda} \chi^{(1)} \nonumber\\
&=& g^{\beta \alpha}\nabla_{\alpha}\Delta \chi^{(1)} + K  g^{\beta \alpha} \nabla_{\alpha} \chi^{(1)} 
\end{eqnarray}
Here, we have used the identity Eq.~\ref{Eq:App_Riemann_4}.

Using the same operations, it is
\begin{eqnarray}
&& \sqrt{g}\nabla_{\alpha}(g^{\alpha\beta}\Delta \chi^{(1)})\nabla_{\beta}\chi^{(1)} \nonumber \\
&=& \partial_{\alpha}\left(\sqrt{g}\Delta  \chi^{(1)} g^{\alpha\beta}\nabla_{\beta} \chi^{(1)} \right)-\sqrt{g}(\Delta \chi^{(1)})^2 \ .
\end{eqnarray}
Hence, the second term in Eq.~\ref{Eq:App_tf_form_linear_free_energy} becomes 
\begin{equation}
-\frac{1+\nu_p}{2Y}\int d^2u\sqrt{g} K g^{\alpha \beta}\nabla_{\alpha} \chi^{(1)}\nabla_{\beta} \chi^{(1)}
\end{equation}
plus a total derivative
\begin{eqnarray}
&& \frac{1+\nu_p}{2Y} \int d^2u\partial_{\alpha}\left[\sqrt{g}\left(T^{\alpha \beta}\nabla_{\beta}-\Delta\chi^{(1)} g^{\alpha\beta}\nabla_{\beta} \right) \chi^{(1)} \right]\nonumber\\
&=&-\frac{1+\nu_p}{2Y} \int d^2u \partial_{\alpha}\left[\sqrt{g}\sigma^{\alpha\beta}\nabla_{\beta}  \chi^{(1)}\right] \ ,
\end{eqnarray}
where use has been made of the definition of the stress tensor, see Eq.~\ref{Eq:form_linear_stress}. The above integral contributes only at the boundary, leading to the contribution
\begin{equation}
-\frac{1+\nu_p}{2Y}\oint dx^{\rho} \sqrt{g}\epsilon_{\rho \alpha} \sigma^{\alpha\beta} \nabla_{\beta}  \chi^{(1)}.
\end{equation}
For a spherical cap, the above equation is
\begin{equation}
\frac{1+\nu_p}{2Y}\oint d\theta \sqrt{g} \sigma^{r \beta} \nabla_{\beta}  \chi^{(1)}.
\end{equation}
and therefore, in the absence of line tension vanishes by the boundary condition $\sigma^{r \beta}=0$, $\beta=r,\theta$ at the boundary.

\section{General Formulas in Riemannian geometry}\label{App_Riemann_Geom}

\subsection{Useful identities}

The following results apply for any metric $g_{\mu \nu}$ in any dimension, unless further restrictions are stated.
\begin{equation}\label{Eq:App_Riemann_0}
\frac{1}{2}\partial_{\mu}(\log{g}) = \Gamma\indices*{^\rho_\mu_\rho} \ .
\end{equation}
The last equation can be written also as
\begin{equation}\label{Eq:App_Riemann_1}
\frac{1}{\sqrt{g}}\partial_{\mu} (\sqrt{g}) = \Gamma\indices*{^\rho_\mu_\rho} \ .
\end{equation}
Another relation involving Christoffel symbols is
\begin{equation}\label{Eq:App_Riemann_2}
  g^{\rho \gamma} \Gamma\indices*{^\nu_\rho_\gamma} = -\frac{1}{\sqrt{g}} \partial_{\gamma}(\sqrt{g}g^{\gamma \nu}) \ .
\end{equation}
The following relation, involving the Riemann tensor is
\begin{equation}\label{Eq:App_Riemann_3}
  [\nabla_{\mu}, \nabla_{\nu}] V^{\rho} = R\indices{^\rho_{\lambda}_\mu_\nu} V^{\lambda} \ .
\end{equation}
The same relation exists for forms as well, namely
\begin{equation}\label{Eq:App_Riemann_4}
  [\nabla_{\mu}, \nabla_{\nu}] W_{\rho \gamma} = -R\indices{^\lambda_{\rho}_\mu_\nu} W_{\lambda \gamma} - R\indices{^\lambda_{\gamma}_\mu_\nu} W_{\rho \lambda}
\end{equation}
Finally, the Ricci and scalar curvature are defined as
\begin{equation}\label{Eq:App_Riemann_ricci}
  R_{\mu \nu} = R\indices{^\lambda_{\mu}_\lambda_\nu} \quad \quad R = g^{\mu \nu} R_{\mu \nu}
\end{equation}
The equations from here onwards are valid in two dimensions only:
\begin{equation}\label{Eq:App_Riemann_5X}
  \frac{1}{g} \epsilon^{\alpha \rho}\epsilon^{\mu \nu} = g^{\alpha \mu} g^{\rho\nu} - g^{\alpha\nu} g^{\rho\mu} \ .
\end{equation}
\begin{equation}\label{Eq:App_Riemann_6}
  g^{\alpha\beta} = \frac{1}{g} \epsilon^{\alpha \rho}\epsilon^{\beta \sigma} g_{\rho\sigma} \ .
\end{equation}
And, the Riemann tensor is
\begin{equation}\label{Eq:App_Riemann_7}
  R_{\rho \lambda \mu \nu} = K \left(g_{\rho\mu}g_{\lambda\nu}- g_{\rho\nu}g_{\lambda\mu}\right) \ ,
\end{equation}
where $K=R/2$ is the Gaussian curvature.

\subsection{Expansion around a given metric}

From the incompatibility expansion Eq.~\ref{Eq:form_expansion} it is
\begin{equation}\label{Eq:App_crystoffel}
  {\bar \Gamma}^{\rho}_{\mu \alpha} = \Gamma^{\rho}_{\mu \alpha} + \eta \tensor*[]{\Gamma}{^{\rho}_\mu_\alpha^{(1)}} + \eta^2 \tensor*[]{\Gamma}{^{\rho}_\mu_\alpha^{(2)}} + \cdots
\end{equation}
here, the $\eta$ value is just a formal quantity that allows to keep track of the different orders in the expansion.

The compatibility of the connection with the metric implies
\begin{eqnarray}\label{Eq:App_comp}
  \nabla_{\mu} g_{\alpha \beta} &=& 0 \nonumber \\
  {\bar \nabla}_{\mu} {\bar g}_{\alpha \beta} & = & 0
\end{eqnarray}
This last equation, in explicit terms is
\begin{equation}\label{Eq:App_connection_bar}
 {\bar \nabla}_{\mu} {\bar g}_{\alpha \beta} = \frac{\partial {\bar g}_{\alpha \beta}}{\partial x_{\mu}} - {\bar \Gamma}^{\rho}_{\mu \alpha} {\bar g}_{\rho \beta} - {\bar \Gamma}^{\rho}_{\mu \beta} {\bar g}_{\alpha \rho} = 0.
\end{equation}
Introducing the expansion Eq.~\ref{Eq:App_crystoffel} into the previous equation leads to
\begin{eqnarray}\label{Eq:App_expansion}
  {\nabla}_{\mu} g_{\alpha \beta} + \eta \left(\nabla_{\mu} g\indices*{_\alpha_\beta^{(1)}} - \tensor*[]{\Gamma}{^{\rho}_\mu_\alpha^{(1)}} g_{\rho \beta} -  \tensor*[]{\Gamma}{^{\rho}_\mu_\beta^{(1)}} g_{\alpha \rho} \right) &+& \nonumber \\
   \eta^2 \left(\nabla_{\mu} g\indices*{_\alpha_\beta^{(2)}}
   - \tensor*[]{\Gamma}{^{\rho}_\mu_\alpha^{(2)}} g_{\rho \beta} -  \tensor*[]{\Gamma}{^{\rho}_\mu_\beta^{(2)}} g_{\alpha \rho}
   - \tensor*[]{\Gamma}{^{\rho}_\mu_\alpha^{(1)}} g\indices*{_\rho_\beta^{(1)}} - \tensor*[]{\Gamma}{^{\rho}_\mu_\beta^{(1)}} g\indices*{_\alpha_\rho^{(1)}}
   \right) && \nonumber\\
   &&
\end{eqnarray}
which immediately leads to the identities
\begin{eqnarray}\label{Eq:App_expansion_equations}
   \nabla_{\mu} g\indices*{_\alpha_\beta^{(1)}} - \tensor*[]{\Gamma}{^{\rho}_\mu_\alpha^{(1)}} g_{\rho \beta} -  \tensor*[]{\Gamma}{^{\rho}_\mu_\beta^{(1)}} g_{\alpha \rho} &=& 0  \nonumber \\ \nonumber
   \nabla_{\mu} g\indices*{_\alpha_\beta^{(2)}}
   - \tensor*[]{\Gamma}{^{\rho}_\mu_\alpha^{(2)}} g_{\rho \beta} -  \tensor*[]{\Gamma}{^{\rho}_\mu_\beta^{(2)}} g_{\alpha \rho}
   - \tensor*[]{\Gamma}{^{\rho}_\mu_\alpha^{(1)}} g\indices*{_\rho_\beta^{(1)}} - \tensor*[]{\Gamma}{^{\rho}_\mu_\beta^{(1)}} g\indices*{_\alpha_\rho^{(1)}}
   &=& 0 \\\
   &&
\end{eqnarray}
with solutions
\begin{equation}\label{Eq:App_linear}
  \tensor*[]{\Gamma}{^{\rho}_\mu_\alpha^{(1)}} = \frac{g^{\rho \beta}}{2}\left( \nabla_{\mu} g\indices*{_\alpha_\beta^{(1)}} + \nabla_{\alpha} g\indices*{_\beta_\mu^{(1)}} - \nabla_{\beta} g\indices*{_\mu_\alpha^{(1)}} \right)
\end{equation}
and
\begin{equation}\label{Eq:App_second}
  \tensor*[]{\Gamma}{^{\rho}_\mu_\alpha^{(2)}} = \frac{g^{\rho \beta}}{2}\left( \nabla_{\mu} g\indices*{_\alpha_\beta^{(2)}} + \nabla_{\alpha} g\indices*{_\beta_\mu^{(2)}} - \nabla_{\beta} g\indices*{_\mu_\alpha^{(2)}} \right) - g^{\rho\beta}\tensor*[]{\Gamma}{^{\gamma}_\mu_\alpha^{(1)}} g\indices*{_\gamma_\beta^{(1)}}
\end{equation}
These expressions allow to compute the Riemann tensor, defined from
\begin{equation}\label{Eq:App_Riemann_tensor}
  {\bar R}\indices{^\rho_\sigma_\mu_\nu} = \partial_{\mu} \tensor*[]{\bar \Gamma}{^{\rho}_\nu_\sigma} - \partial_{\nu} \tensor*[]{\bar \Gamma}{^{\rho}_\mu_\sigma} + \tensor*[]{\bar \Gamma}{^{\rho}_\mu_\lambda} \tensor*[]{\bar \Gamma}{^{\lambda}_\nu_\sigma} -
  \tensor*[]{\bar \Gamma}{^{\rho}_\nu_\lambda} \tensor*[]{\bar \Gamma}{^{\lambda}_\mu_\sigma} \ .
\end{equation}
Inserting the terms in Eq.~\ref{Eq:App_linear} and Eq.~\ref{Eq:App_second} after some algebra it leads to
\begin{eqnarray}\label{Eq:App_Riemann_expansion}
{\bar R}\indices{^\rho_\sigma_\mu_\nu} & = & R\indices{^\rho_\sigma_\mu_\nu} + \eta\left(\nabla_{\mu}\tensor*[]{\Gamma}{^{\rho}_\nu_\sigma^{(1)}} - \nabla_{\nu}\tensor*[]{\Gamma}{^{\rho}_\mu_\sigma^{(1)}}\right) + \\\nonumber
&+& \eta^2\left(\nabla_{\mu}\tensor*[]{\Gamma}{^{\rho}_\nu_\sigma^{(2)}} - \nabla_{\nu}\tensor*[]{\Gamma}{^{\rho}_\mu_\sigma^{(2)}} +  \tensor*[]{\Gamma}{^{\rho}_\mu_\lambda^{(1)}}\tensor*[]{\Gamma}{^{\lambda}_\nu_\sigma^{(1)}} - \tensor*[]{\Gamma}{^{\rho}_\nu_\lambda^{(1)}}\tensor*[]{\Gamma}{^{\lambda}_\mu_\sigma^{(1)}}  \right) \ .
\end{eqnarray}
The Ricci tensor is
\begin{eqnarray}\label{Eq:App_Riemann_expansion_ricci}
{\bar R}_{\sigma \nu} & = & R_{\sigma \nu} + \eta\left(\nabla_{\mu}\tensor*[]{\Gamma}{^{\mu}_\nu_\sigma^{(1)}} - \nabla_{\nu}\tensor*[]{\Gamma}{^{\mu}_\mu_\sigma^{(1)}}\right) + \\\nonumber
&+& \eta^2\left(\nabla_{\mu}\tensor*[]{\Gamma}{^{\mu}_\nu_\sigma^{(2)}} - \nabla_{\nu}\tensor*[]{\Gamma}{^{\mu}_\mu_\sigma^{(2)}} +  \tensor*[]{\Gamma}{^{\mu}_\mu_\lambda^{(1)}}\tensor*[]{\Gamma}{^{\lambda}_\nu_\sigma^{(1)}} - \tensor*[]{\Gamma}{^{\mu}_\nu_\lambda^{(1)}}\tensor*[]{\Gamma}{^{\lambda}_\mu_\sigma^{(1)}}  \right) \ .
\end{eqnarray}
\bibliography{GeneralReferences,Mine}

\begin{thebibliography}{41}%
\makeatletter
\providecommand \@ifxundefined [1]{%
 \@ifx{#1\undefined}
}%
\providecommand \@ifnum [1]{%
 \ifnum #1\expandafter \@firstoftwo
 \else \expandafter \@secondoftwo
 \fi
}%
\providecommand \@ifx [1]{%
 \ifx #1\expandafter \@firstoftwo
 \else \expandafter \@secondoftwo
 \fi
}%
\providecommand \natexlab [1]{#1}%
\providecommand \enquote  [1]{``#1''}%
\providecommand \bibnamefont  [1]{#1}%
\providecommand \bibfnamefont [1]{#1}%
\providecommand \citenamefont [1]{#1}%
\providecommand \href@noop [0]{\@secondoftwo}%
\providecommand \href [0]{\begingroup \@sanitize@url \@href}%
\providecommand \@href[1]{\@@startlink{#1}\@@href}%
\providecommand \@@href[1]{\endgroup#1\@@endlink}%
\providecommand \@sanitize@url [0]{\catcode `\\12\catcode `\$12\catcode
  `\&12\catcode `\#12\catcode `\^12\catcode `\_12\catcode `\%12\relax}%
\providecommand \@@startlink[1]{}%
\providecommand \@@endlink[0]{}%
\providecommand \url  [0]{\begingroup\@sanitize@url \@url }%
\providecommand \@url [1]{\endgroup\@href {#1}{\urlprefix }}%
\providecommand \urlprefix  [0]{URL }%
\providecommand \Eprint [0]{\href }%
\providecommand \doibase [0]{http://dx.doi.org/}%
\providecommand \selectlanguage [0]{\@gobble}%
\providecommand \bibinfo  [0]{\@secondoftwo}%
\providecommand \bibfield  [0]{\@secondoftwo}%
\providecommand \translation [1]{[#1]}%
\providecommand \BibitemOpen [0]{}%
\providecommand \bibitemStop [0]{}%
\providecommand \bibitemNoStop [0]{.\EOS\space}%
\providecommand \EOS [0]{\spacefactor3000\relax}%
\providecommand \BibitemShut  [1]{\csname bibitem#1\endcsname}%
\let\auto@bib@innerbib\@empty
\bibitem [{\citenamefont {Bausch}\ \emph {et~al.}(2003)\citenamefont {Bausch},
  \citenamefont {Bowick}, \citenamefont {Cacciuto}, \citenamefont {Dinsmore},
  \citenamefont {Hsu}, \citenamefont {Nelson}, \citenamefont {Nikolaides},
  \citenamefont {Travesset},\ and\ \citenamefont {Weitz}}]{BauschMe2003}%
  \BibitemOpen
  \bibfield  {author} {\bibinfo {author} {\bibfnamefont {A.~R.}\ \bibnamefont
  {Bausch}}, \bibinfo {author} {\bibfnamefont {M.~J.}\ \bibnamefont {Bowick}},
  \bibinfo {author} {\bibfnamefont {A.}~\bibnamefont {Cacciuto}}, \bibinfo
  {author} {\bibfnamefont {A.~D.}\ \bibnamefont {Dinsmore}}, \bibinfo {author}
  {\bibfnamefont {M.~F.}\ \bibnamefont {Hsu}}, \bibinfo {author} {\bibfnamefont
  {D.~R.}\ \bibnamefont {Nelson}}, \bibinfo {author} {\bibfnamefont {M.~G.}\
  \bibnamefont {Nikolaides}}, \bibinfo {author} {\bibfnamefont
  {A.}~\bibnamefont {Travesset}}, \ and\ \bibinfo {author} {\bibfnamefont
  {D.~A.}\ \bibnamefont {Weitz}},\ }\href
  {http://www.sciencemag.org/content/298/5595/1006.full} {\bibfield  {journal}
  {\bibinfo  {journal} {Science}\ }\textbf {\bibinfo {volume} {299}},\ \bibinfo
  {pages} {1716} (\bibinfo {year} {2003})}\BibitemShut {NoStop}%
\bibitem [{\citenamefont {Panahandeh}\ \emph {et~al.}(2018)\citenamefont
  {Panahandeh}, \citenamefont {Li},\ and\ \citenamefont {Zandi}}]{Sanaz2018}%
  \BibitemOpen
  \bibfield  {author} {\bibinfo {author} {\bibfnamefont {S.}~\bibnamefont
  {Panahandeh}}, \bibinfo {author} {\bibfnamefont {S.}~\bibnamefont {Li}}, \
  and\ \bibinfo {author} {\bibfnamefont {R.}~\bibnamefont {Zandi}},\ }\href
  {\doibase 10.1039/C8NR07202G} {\bibfield  {journal} {\bibinfo  {journal}
  {Nanoscale}\ }\textbf {\bibinfo {volume} {10}},\ \bibinfo {pages} {22802}
  (\bibinfo {year} {2018})}\BibitemShut {NoStop}%
\bibitem [{\citenamefont {Irvine}\ \emph {et~al.}(2010)\citenamefont {Irvine},
  \citenamefont {Vitelli},\ and\ \citenamefont
  {Chaikin}}]{IrvineVitelliChaikin2010}%
  \BibitemOpen
  \bibfield  {author} {\bibinfo {author} {\bibfnamefont {W.~T.~M.}\
  \bibnamefont {Irvine}}, \bibinfo {author} {\bibfnamefont {V.}~\bibnamefont
  {Vitelli}}, \ and\ \bibinfo {author} {\bibfnamefont {P.~M.}\ \bibnamefont
  {Chaikin}},\ }\href {http://dx.doi.org/10.1038/nature09620} {\bibfield
  {journal} {\bibinfo  {journal} {Nature}\ }\textbf {\bibinfo {volume} {468}},\
  \bibinfo {pages} {947} (\bibinfo {year} {2010})}\BibitemShut {NoStop}%
\bibitem [{\citenamefont {Lidmar}\ \emph {et~al.}(2003)\citenamefont {Lidmar},
  \citenamefont {Mirny},\ and\ \citenamefont {Nelson}}]{Lidmar2003}%
  \BibitemOpen
  \bibfield  {author} {\bibinfo {author} {\bibfnamefont {J.}~\bibnamefont
  {Lidmar}}, \bibinfo {author} {\bibfnamefont {L.}~\bibnamefont {Mirny}}, \
  and\ \bibinfo {author} {\bibfnamefont {D.~R.}\ \bibnamefont {Nelson}},\
  }\href {https://link.aps.org/doi/10.1103/PhysRevE.68.051910} {\bibfield
  {journal} {\bibinfo  {journal} {Phys. Rev. E}\ }\textbf {\bibinfo {volume}
  {68}},\ \bibinfo {pages} {051910} (\bibinfo {year} {2003})}\BibitemShut
  {NoStop}%
\bibitem [{\citenamefont {Wagner}\ and\ \citenamefont
  {Zandi}(2015)}]{Wagner2015}%
  \BibitemOpen
  \bibfield  {author} {\bibinfo {author} {\bibfnamefont {J.}~\bibnamefont
  {Wagner}}\ and\ \bibinfo {author} {\bibfnamefont {R.}~\bibnamefont {Zandi}},\
  }\href@noop {} {\bibfield  {journal} {\bibinfo  {journal} {Biophysical
  journal}\ }\textbf {\bibinfo {volume} {109}},\ \bibinfo {pages} {956}
  (\bibinfo {year} {2015})}\BibitemShut {NoStop}%
\bibitem [{\citenamefont {Ning}\ \emph {et~al.}(2016)\citenamefont {Ning},
  \citenamefont {Erdemci-Tandogan}, \citenamefont {Yufenyuy}, \citenamefont
  {Wagner}, \citenamefont {Himes}, \citenamefont {Zhao}, \citenamefont {Aiken},
  \citenamefont {Zandi},\ and\ \citenamefont {Zhang}}]{ning2016vitro}%
  \BibitemOpen
  \bibfield  {author} {\bibinfo {author} {\bibfnamefont {J.}~\bibnamefont
  {Ning}}, \bibinfo {author} {\bibfnamefont {G.}~\bibnamefont
  {Erdemci-Tandogan}}, \bibinfo {author} {\bibfnamefont {E.~L.}\ \bibnamefont
  {Yufenyuy}}, \bibinfo {author} {\bibfnamefont {J.}~\bibnamefont {Wagner}},
  \bibinfo {author} {\bibfnamefont {B.~A.}\ \bibnamefont {Himes}}, \bibinfo
  {author} {\bibfnamefont {G.}~\bibnamefont {Zhao}}, \bibinfo {author}
  {\bibfnamefont {C.}~\bibnamefont {Aiken}}, \bibinfo {author} {\bibfnamefont
  {R.}~\bibnamefont {Zandi}}, \ and\ \bibinfo {author} {\bibfnamefont
  {P.}~\bibnamefont {Zhang}},\ }\href@noop {} {\bibfield  {journal} {\bibinfo
  {journal} {Nature communications}\ }\textbf {\bibinfo {volume} {7}},\
  \bibinfo {pages} {13689} (\bibinfo {year} {2016})}\BibitemShut {NoStop}%
\bibitem [{\citenamefont {Vernizzi}\ and\ \citenamefont {Olvera de~la
  Cruz}(2007)}]{VernizziOlveradelaCruz2007}%
  \BibitemOpen
  \bibfield  {author} {\bibinfo {author} {\bibfnamefont {G.}~\bibnamefont
  {Vernizzi}}\ and\ \bibinfo {author} {\bibfnamefont {M.}~\bibnamefont {Olvera
  de~la Cruz}},\ }\href {http://www.pnas.org/content/104/47/18382.abstract N2 -
  Shells of various viruses and other closed packed structures with spherical
  topology exhibit icosahedral symmetry because the surface of a sphere cannot
  be tiled without defects, and icosahedral symmetry yields the most symmetric
  configuration with the minimum number of defects. Icosahedral symmetry is
  different from icosahedral-shaped structures, which include some large
  viruses, cationic-anionic vesicles, and fullerenes. We present a faceting
  mechanism of ionic shells into icosahedral shapes that breaks icosahedral
  symmetry resulting from different arrangements of the charged components
  among the facets. These self-organized ionic structures may favor the
  formation of flat domains on curved surfaces. We show that icosahedral shapes
  without rotational symmetry can have lower energy than spheres with
  icosahedral symmetry caused by preferred bending directions in the planar
  ionic lattice. The ability to create icosahedral shapes without icosahedral
  symmetry may lead to the design of new functional materials. The
  electrostatically driven faceting mechanism we present here suggests that we
  can design faceted polyhedra with diverse symmetries by coassembling
  oppositely charged molecules of different stoichiometric ratios.} {\bibfield
  {journal} {\bibinfo  {journal} {Proceedings of the National Academy of
  Sciences}\ }\textbf {\bibinfo {volume} {104}},\ \bibinfo {pages} {18382}
  (\bibinfo {year} {2007})}\BibitemShut {NoStop}%
\bibitem [{\citenamefont {Landau}\ and\ \citenamefont
  {Lifshitz}(1985)}]{LandauElasticityBook}%
  \BibitemOpen
  \bibfield  {author} {\bibinfo {author} {\bibfnamefont {L.}~\bibnamefont
  {Landau}}\ and\ \bibinfo {author} {\bibfnamefont {E.}~\bibnamefont
  {Lifshitz}},\ }\href@noop {} {\emph {\bibinfo {title} {Theory of
  Elasticity}}}\ (\bibinfo  {publisher} {Butterworth-Heinemann; 3 edition},\
  \bibinfo {year} {1985})\BibitemShut {NoStop}%
\bibitem [{\citenamefont {Klein}\ \emph {et~al.}(2007)\citenamefont {Klein},
  \citenamefont {Efrati},\ and\ \citenamefont {Sharon}}]{Klein1116}%
  \BibitemOpen
  \bibfield  {author} {\bibinfo {author} {\bibfnamefont {Y.}~\bibnamefont
  {Klein}}, \bibinfo {author} {\bibfnamefont {E.}~\bibnamefont {Efrati}}, \
  and\ \bibinfo {author} {\bibfnamefont {E.}~\bibnamefont {Sharon}},\ }\href
  {\doibase 10.1126/science.1135994} {\bibfield  {journal} {\bibinfo  {journal}
  {Science}\ }\textbf {\bibinfo {volume} {315}},\ \bibinfo {pages} {1116}
  (\bibinfo {year} {2007})}\BibitemShut {NoStop}%
\bibitem [{\citenamefont {Davidovitch}\ \emph {et~al.}(2018)\citenamefont
  {Davidovitch}, \citenamefont {Sun},\ and\ \citenamefont
  {Grason}}]{Davidovitvh2019}%
  \BibitemOpen
  \bibfield  {author} {\bibinfo {author} {\bibfnamefont {B.}~\bibnamefont
  {Davidovitch}}, \bibinfo {author} {\bibfnamefont {Y.}~\bibnamefont {Sun}}, \
  and\ \bibinfo {author} {\bibfnamefont {G.~M.}\ \bibnamefont {Grason}},\
  }\href {\doibase 10.1073/pnas.1815507116} {\bibfield  {journal} {\bibinfo
  {journal} {Proceedings of the National Academy of Sciences}\ }\textbf
  {\bibinfo {volume} {116}},\ \bibinfo {pages} {1483} (\bibinfo {year}
  {2018})},\ \Eprint {http://arxiv.org/abs/1809.06919} {arXiv:1809.06919}
  \BibitemShut {NoStop}%
\bibitem [{\citenamefont {Sadoc}\ and\ \citenamefont
  {Mosseri}(1999)}]{SadocBook1999}%
  \BibitemOpen
  \bibfield  {author} {\bibinfo {author} {\bibfnamefont {J.}~\bibnamefont
  {Sadoc}}\ and\ \bibinfo {author} {\bibfnamefont {R.}~\bibnamefont
  {Mosseri}},\ }\href@noop {} {\emph {\bibinfo {title} {Geometrical
  Frustration}}}\ (\bibinfo  {publisher} {Cambridge University Press},\
  \bibinfo {year} {1999})\BibitemShut {NoStop}%
\bibitem [{\citenamefont {Nelson}(1983)}]{Nelson1983}%
  \BibitemOpen
  \bibfield  {author} {\bibinfo {author} {\bibfnamefont {D.~R.}\ \bibnamefont
  {Nelson}},\ }\href {https://link.aps.org/doi/10.1103/PhysRevB.28.5515}
  {\bibfield  {journal} {\bibinfo  {journal} {Phys. Rev. B}\ }\textbf {\bibinfo
  {volume} {28}},\ \bibinfo {pages} {5515} (\bibinfo {year}
  {1983})}\BibitemShut {NoStop}%
\bibitem [{\citenamefont {Travesset}(2017)}]{Travesset2017b}%
  \BibitemOpen
  \bibfield  {author} {\bibinfo {author} {\bibfnamefont {A.}~\bibnamefont
  {Travesset}},\ }\href
  {https://link.aps.org/doi/10.1103/PhysRevLett.119.115701} {\bibfield
  {journal} {\bibinfo  {journal} {Phys. Rev. Lett.}\ }\textbf {\bibinfo
  {volume} {119}},\ \bibinfo {pages} {115701} (\bibinfo {year}
  {2017})}\BibitemShut {NoStop}%
\bibitem [{\citenamefont {Nelson}(2002)}]{NelsonBook2002}%
  \BibitemOpen
  \bibfield  {author} {\bibinfo {author} {\bibfnamefont {D.}~\bibnamefont
  {Nelson}},\ }\href@noop {} {\emph {\bibinfo {title} {Defects and Geometry in
  Condensed Matter Physics}}}\ (\bibinfo  {publisher} {Cambridge Press},\
  \bibinfo {year} {2002})\BibitemShut {NoStop}%
\bibitem [{\citenamefont {Chaikin}\ and\ \citenamefont
  {Lubensky}(2003)}]{ChaikinBook2003}%
  \BibitemOpen
  \bibfield  {author} {\bibinfo {author} {\bibfnamefont {P.}~\bibnamefont
  {Chaikin}}\ and\ \bibinfo {author} {\bibfnamefont {T.}~\bibnamefont
  {Lubensky}},\ }\href@noop {} {\emph {\bibinfo {title} {Principles of
  condensed matter physics}}}\ (\bibinfo  {publisher} {Cambridge University
  Press},\ \bibinfo {year} {2003})\BibitemShut {NoStop}%
\bibitem [{\citenamefont {Nakahara}(1990)}]{Nakaharabook1990}%
  \BibitemOpen
  \bibfield  {author} {\bibinfo {author} {\bibfnamefont {M.}~\bibnamefont
  {Nakahara}},\ }\href@noop {} {\emph {\bibinfo {title} {Geometry, Topology and
  Physics}}}\ (\bibinfo  {publisher} {Adam Hilger},\ \bibinfo {year}
  {1990})\BibitemShut {NoStop}%
\bibitem [{\citenamefont {Seung}\ and\ \citenamefont
  {Nelson}(1988)}]{Seung1988}%
  \BibitemOpen
  \bibfield  {author} {\bibinfo {author} {\bibfnamefont {H.~S.}\ \bibnamefont
  {Seung}}\ and\ \bibinfo {author} {\bibfnamefont {D.~R.}\ \bibnamefont
  {Nelson}},\ }\href {\doibase 10.1103/PhysRevA.38.1005} {\bibfield  {journal}
  {\bibinfo  {journal} {Phys. Rev. A}\ }\textbf {\bibinfo {volume} {38}},\
  \bibinfo {pages} {1005} (\bibinfo {year} {1988})}\BibitemShut {NoStop}%
\bibitem [{\citenamefont {Schneider}\ and\ \citenamefont
  {Gompper}(2005)}]{Schneider2005}%
  \BibitemOpen
  \bibfield  {author} {\bibinfo {author} {\bibfnamefont {S.}~\bibnamefont
  {Schneider}}\ and\ \bibinfo {author} {\bibfnamefont {G.}~\bibnamefont
  {Gompper}},\ }\href@noop {} {\bibfield  {journal} {\bibinfo  {journal}
  {Europhys. Lett.}\ }\textbf {\bibinfo {volume} {70}},\ \bibinfo {pages} {136}
  (\bibinfo {year} {2005})}\BibitemShut {NoStop}%
\bibitem [{\citenamefont {Morozov}\ and\ \citenamefont
  {Bruinsma}(2010)}]{MorozovBruinsma2010}%
  \BibitemOpen
  \bibfield  {author} {\bibinfo {author} {\bibfnamefont {A.~Y.}\ \bibnamefont
  {Morozov}}\ and\ \bibinfo {author} {\bibfnamefont {R.~F.}\ \bibnamefont
  {Bruinsma}},\ }\href {\doibase 10.1103/PhysRevE.81.041925} {\bibfield
  {journal} {\bibinfo  {journal} {Phys. Rev. E}\ }\textbf {\bibinfo {volume}
  {81}},\ \bibinfo {pages} {041925} (\bibinfo {year} {2010})}\BibitemShut
  {NoStop}%
\bibitem [{\citenamefont {Grason}(2010)}]{Grason2010}%
  \BibitemOpen
  \bibfield  {author} {\bibinfo {author} {\bibfnamefont {G.~M.}\ \bibnamefont
  {Grason}},\ }\href {https://link.aps.org/doi/10.1103/PhysRevLett.105.045502}
  {\bibfield  {journal} {\bibinfo  {journal} {Phys. Rev. Lett.}\ }\textbf
  {\bibinfo {volume} {105}},\ \bibinfo {pages} {045502} (\bibinfo {year}
  {2010})}\BibitemShut {NoStop}%
\bibitem [{\citenamefont {Grason}(2012)}]{Grason2012}%
  \BibitemOpen
  \bibfield  {author} {\bibinfo {author} {\bibfnamefont {G.~M.}\ \bibnamefont
  {Grason}},\ }\href {https://link.aps.org/doi/10.1103/PhysRevE.85.031603}
  {\bibfield  {journal} {\bibinfo  {journal} {Phys. Rev. E}\ }\textbf {\bibinfo
  {volume} {85}},\ \bibinfo {pages} {031603} (\bibinfo {year}
  {2012})}\BibitemShut {NoStop}%
\bibitem [{\citenamefont {Grason}(2015)}]{Grason2015}%
  \BibitemOpen
  \bibfield  {author} {\bibinfo {author} {\bibfnamefont {G.~M.}\ \bibnamefont
  {Grason}},\ }\href {https://link.aps.org/doi/10.1103/RevModPhys.87.401}
  {\bibfield  {journal} {\bibinfo  {journal} {Reviews of Modern Physics}\
  }\textbf {\bibinfo {volume} {87}},\ \bibinfo {pages} {401} (\bibinfo {year}
  {2015})}\BibitemShut {NoStop}%
\bibitem [{\citenamefont {Castelnovo}(2017)}]{Castelnovo2017}%
  \BibitemOpen
  \bibfield  {author} {\bibinfo {author} {\bibfnamefont {M.}~\bibnamefont
  {Castelnovo}},\ }\href {\doibase 10.1103/PhysRevE.95.052405} {\bibfield
  {journal} {\bibinfo  {journal} {Phys. Rev. E}\ }\textbf {\bibinfo {volume}
  {95}},\ \bibinfo {pages} {052405} (\bibinfo {year} {2017})}\BibitemShut
  {NoStop}%
\bibitem [{\citenamefont {Bowick}\ \emph {et~al.}(2000)\citenamefont {Bowick},
  \citenamefont {Nelson},\ and\ \citenamefont {Travesset}}]{BowickMe2000}%
  \BibitemOpen
  \bibfield  {author} {\bibinfo {author} {\bibfnamefont {M.~J.}\ \bibnamefont
  {Bowick}}, \bibinfo {author} {\bibfnamefont {D.~R.}\ \bibnamefont {Nelson}},
  \ and\ \bibinfo {author} {\bibfnamefont {A.}~\bibnamefont {Travesset}},\
  }\href {http://link.aps.org/doi/10.1103/PhysRevB.62.8738} {\bibfield
  {journal} {\bibinfo  {journal} {Physical Review B}\ }\textbf {\bibinfo
  {volume} {62}},\ \bibinfo {pages} {8738} (\bibinfo {year}
  {2000})}\BibitemShut {NoStop}%
\bibitem [{\citenamefont {Bowick}\ \emph {et~al.}(2002)\citenamefont {Bowick},
  \citenamefont {Cacciuto}, \citenamefont {Nelson},\ and\ \citenamefont
  {Travesset}}]{BowickMe2002}%
  \BibitemOpen
  \bibfield  {author} {\bibinfo {author} {\bibfnamefont {M.}~\bibnamefont
  {Bowick}}, \bibinfo {author} {\bibfnamefont {A.}~\bibnamefont {Cacciuto}},
  \bibinfo {author} {\bibfnamefont {D.~R.}\ \bibnamefont {Nelson}}, \ and\
  \bibinfo {author} {\bibfnamefont {A.}~\bibnamefont {Travesset}},\ }\href
  {http://link.aps.org/doi/10.1103/PhysRevLett.89.185502} {\bibfield  {journal}
  {\bibinfo  {journal} {Phys. Rev. Lett.}\ }\textbf {\bibinfo {volume} {89}},\
  \bibinfo {pages} {185502} (\bibinfo {year} {2002})}\BibitemShut {NoStop}%
\bibitem [{\citenamefont {Bowick}\ \emph {et~al.}(2006)\citenamefont {Bowick},
  \citenamefont {Cacciuto}, \citenamefont {Nelson},\ and\ \citenamefont
  {Travesset}}]{BowickMea2006}%
  \BibitemOpen
  \bibfield  {author} {\bibinfo {author} {\bibfnamefont {M.~J.}\ \bibnamefont
  {Bowick}}, \bibinfo {author} {\bibfnamefont {A.}~\bibnamefont {Cacciuto}},
  \bibinfo {author} {\bibfnamefont {D.~R.}\ \bibnamefont {Nelson}}, \ and\
  \bibinfo {author} {\bibfnamefont {A.}~\bibnamefont {Travesset}},\ }\href
  {http://link.aps.org/doi/10.1103/PhysRevLett.89.185502} {\bibfield  {journal}
  {\bibinfo  {journal} {Phys. Rev. B}\ }\textbf {\bibinfo {volume} {73}},\
  \bibinfo {pages} {024115} (\bibinfo {year} {2006})}\BibitemShut {NoStop}%
\bibitem [{\citenamefont {Giomi}\ and\ \citenamefont
  {Bowick}(2007)}]{GiomiBowick2007}%
  \BibitemOpen
  \bibfield  {author} {\bibinfo {author} {\bibfnamefont {L.}~\bibnamefont
  {Giomi}}\ and\ \bibinfo {author} {\bibfnamefont {M.}~\bibnamefont {Bowick}},\
  }\href {\doibase 10.1103/PhysRevB.76.054106} {\bibfield  {journal} {\bibinfo
  {journal} {Phys. Rev. B}\ }\textbf {\bibinfo {volume} {76}},\ \bibinfo
  {pages} {054106} (\bibinfo {year} {2007})}\BibitemShut {NoStop}%
\bibitem [{\citenamefont {Kondo}(1955)}]{Kondo1955}%
  \BibitemOpen
  \bibfield  {author} {\bibinfo {author} {\bibfnamefont {K.}~\bibnamefont
  {Kondo}},\ }\href@noop {} {\emph {\bibinfo {title} {Geometry of elastic
  deformation and incompatibility, RAAG Memoirs, Volume 1, Division C}}}\
  (\bibinfo  {publisher} {Gakujutsu Bunken Fukuy-kai},\ \bibinfo {address}
  {Tokyio},\ \bibinfo {year} {1955})\BibitemShut {NoStop}%
\bibitem [{\citenamefont {Koiter}(1966)}]{Koiter1966}%
  \BibitemOpen
  \bibfield  {author} {\bibinfo {author} {\bibfnamefont {W.}~\bibnamefont
  {Koiter}},\ }\href@noop {} {\bibfield  {journal} {\bibinfo  {journal} {Proc.
  Kon. Ned. Acad. Wetensch. B}\ }\textbf {\bibinfo {volume} {69}},\ \bibinfo
  {pages} {1} (\bibinfo {year} {1966})}\BibitemShut {NoStop}%
\bibitem [{\citenamefont {Efrati}\ \emph {et~al.}(2009)\citenamefont {Efrati},
  \citenamefont {Sharon},\ and\ \citenamefont {Kupferman}}]{Efrati2009}%
  \BibitemOpen
  \bibfield  {author} {\bibinfo {author} {\bibfnamefont {E.}~\bibnamefont
  {Efrati}}, \bibinfo {author} {\bibfnamefont {E.}~\bibnamefont {Sharon}}, \
  and\ \bibinfo {author} {\bibfnamefont {R.}~\bibnamefont {Kupferman}},\ }\href
  {http://www.sciencedirect.com/science/article/pii/S0022509608002160}
  {\bibfield  {journal} {\bibinfo  {journal} {Journal of the Mechanics and
  Physics of Solids}\ }\textbf {\bibinfo {volume} {57}},\ \bibinfo {pages}
  {762} (\bibinfo {year} {2009})}\BibitemShut {NoStop}%
\bibitem [{\citenamefont {Moshe}\ \emph {et~al.}(2015)\citenamefont {Moshe},
  \citenamefont {Sharon},\ and\ \citenamefont
  {Kupferman}}]{MosheSharonKupferman2015}%
  \BibitemOpen
  \bibfield  {author} {\bibinfo {author} {\bibfnamefont {M.}~\bibnamefont
  {Moshe}}, \bibinfo {author} {\bibfnamefont {E.}~\bibnamefont {Sharon}}, \
  and\ \bibinfo {author} {\bibfnamefont {R.}~\bibnamefont {Kupferman}},\ }\href
  {\doibase 10.1103/PhysRevE.92.062403} {\bibfield  {journal} {\bibinfo
  {journal} {Phys. Rev. E}\ }\textbf {\bibinfo {volume} {92}},\ \bibinfo
  {pages} {062403} (\bibinfo {year} {2015})}\BibitemShut {NoStop}%
\bibitem [{\citenamefont {Travesset}(2016)}]{Travesset2016a}%
  \BibitemOpen
  \bibfield  {author} {\bibinfo {author} {\bibfnamefont {A.}~\bibnamefont
  {Travesset}},\ }\href {\doibase 10.1103/PhysRevE.94.063001} {\bibfield
  {journal} {\bibinfo  {journal} {Phys. Rev. E}\ }\textbf {\bibinfo {volume}
  {94}},\ \bibinfo {pages} {063001} (\bibinfo {year} {2016})}\BibitemShut
  {NoStop}%
\bibitem [{\citenamefont {Li}\ \emph {et~al.}(2018)\citenamefont {Li},
  \citenamefont {Roy}, \citenamefont {Travesset},\ and\ \citenamefont
  {Zandi}}]{LiMe2018}%
  \BibitemOpen
  \bibfield  {author} {\bibinfo {author} {\bibfnamefont {S.}~\bibnamefont
  {Li}}, \bibinfo {author} {\bibfnamefont {P.}~\bibnamefont {Roy}}, \bibinfo
  {author} {\bibfnamefont {A.}~\bibnamefont {Travesset}}, \ and\ \bibinfo
  {author} {\bibfnamefont {R.}~\bibnamefont {Zandi}},\ }\href
  {http://www.pnas.org/content/115/43/10971.abstract} {\bibfield  {journal}
  {\bibinfo  {journal} {Proceedings of the National Academy of Sciences}\
  }\textbf {\bibinfo {volume} {115}},\ \bibinfo {pages} {10971} (\bibinfo
  {year} {2018})}\BibitemShut {NoStop}%
\bibitem [{\citenamefont {Efrati}\ \emph {et~al.}(2013)\citenamefont {Efrati},
  \citenamefont {Sharon},\ and\ \citenamefont {Kupferman}}]{Efrati2013}%
  \BibitemOpen
  \bibfield  {author} {\bibinfo {author} {\bibfnamefont {E.}~\bibnamefont
  {Efrati}}, \bibinfo {author} {\bibfnamefont {E.}~\bibnamefont {Sharon}}, \
  and\ \bibinfo {author} {\bibfnamefont {R.}~\bibnamefont {Kupferman}},\ }\href
  {\doibase 10.1039/C3SM50660F} {\bibfield  {journal} {\bibinfo  {journal}
  {Soft Matter}\ }\textbf {\bibinfo {volume} {9}},\ \bibinfo {pages} {8187}
  (\bibinfo {year} {2013})}\BibitemShut {NoStop}%
\bibitem [{\citenamefont {Guven}\ \emph {et~al.}(2013)\citenamefont {Guven},
  \citenamefont {Hanna}, \citenamefont {Kahraman},\ and\ \citenamefont
  {Müller}}]{Guven2013}%
  \BibitemOpen
  \bibfield  {author} {\bibinfo {author} {\bibfnamefont {J.}~\bibnamefont
  {Guven}}, \bibinfo {author} {\bibfnamefont {J.~A.}\ \bibnamefont {Hanna}},
  \bibinfo {author} {\bibfnamefont {O.}~\bibnamefont {Kahraman}}, \ and\
  \bibinfo {author} {\bibfnamefont {M.~M.}\ \bibnamefont {Müller}},\ }\href
  {https://doi.org/10.1140/epje/i2013-13106-0} {\bibfield  {journal} {\bibinfo
  {journal} {Eur. Phys. J. E}\ }\textbf {\bibinfo {volume} {36}} (\bibinfo
  {year} {2013})}\BibitemShut {NoStop}%
\bibitem [{\citenamefont {Bowick}\ and\ \citenamefont
  {Travesset}(2001)}]{BowickMed2001}%
  \BibitemOpen
  \bibfield  {author} {\bibinfo {author} {\bibfnamefont {M.}~\bibnamefont
  {Bowick}}\ and\ \bibinfo {author} {\bibfnamefont {A.}~\bibnamefont
  {Travesset}},\ }\href {http://dx.doi.org/10.1088/0305-4470/34/8/301}
  {\bibfield  {journal} {\bibinfo  {journal} {J. Phys. A:Math. Gen.}\ }\textbf
  {\bibinfo {volume} {34}},\ \bibinfo {pages} {1535} (\bibinfo {year}
  {2001})}\BibitemShut {NoStop}%
\bibitem [{\citenamefont {Azadi}\ and\ \citenamefont
  {Grason}(2014)}]{AzadiGrason2014}%
  \BibitemOpen
  \bibfield  {author} {\bibinfo {author} {\bibfnamefont {A.}~\bibnamefont
  {Azadi}}\ and\ \bibinfo {author} {\bibfnamefont {G.~M.}\ \bibnamefont
  {Grason}},\ }\href {\doibase 10.1103/PhysRevLett.112.225502} {\bibfield
  {journal} {\bibinfo  {journal} {Phys. Rev. Lett.}\ }\textbf {\bibinfo
  {volume} {112}},\ \bibinfo {pages} {225502} (\bibinfo {year}
  {2014})}\BibitemShut {NoStop}%
\bibitem [{\citenamefont {Azadi}\ and\ \citenamefont
  {Grason}(2016)}]{Azadi2016}%
  \BibitemOpen
  \bibfield  {author} {\bibinfo {author} {\bibfnamefont {A.}~\bibnamefont
  {Azadi}}\ and\ \bibinfo {author} {\bibfnamefont {G.~M.}\ \bibnamefont
  {Grason}},\ }\href {\doibase 10.1103/PhysRevE.94.013003} {\bibfield
  {journal} {\bibinfo  {journal} {Phys. Rev. E}\ }\textbf {\bibinfo {volume}
  {94}},\ \bibinfo {pages} {013003} (\bibinfo {year} {2016})}\BibitemShut
  {NoStop}%
\bibitem [{\citenamefont {Travesset}(2003)}]{TravessetMe2003}%
  \BibitemOpen
  \bibfield  {author} {\bibinfo {author} {\bibfnamefont {A.}~\bibnamefont
  {Travesset}},\ }\href {http://link.aps.org/doi/10.1103/PhysRevB.68.115421}
  {\bibfield  {journal} {\bibinfo  {journal} {Phys. Rev. B}\ }\textbf {\bibinfo
  {volume} {68}},\ \bibinfo {pages} {115421} (\bibinfo {year}
  {2003})}\BibitemShut {NoStop}%
\bibitem [{\citenamefont {Talapin}\ \emph {et~al.}(2009)\citenamefont
  {Talapin}, \citenamefont {Shevchenko}, \citenamefont {Bodnarchuk},
  \citenamefont {Ye}, \citenamefont {Chen},\ and\ \citenamefont
  {Murray}}]{Talapin2009a}%
  \BibitemOpen
  \bibfield  {author} {\bibinfo {author} {\bibfnamefont {D.~V.}\ \bibnamefont
  {Talapin}}, \bibinfo {author} {\bibfnamefont {E.~V.}\ \bibnamefont
  {Shevchenko}}, \bibinfo {author} {\bibfnamefont {M.~I.}\ \bibnamefont
  {Bodnarchuk}}, \bibinfo {author} {\bibfnamefont {X.}~\bibnamefont {Ye}},
  \bibinfo {author} {\bibfnamefont {J.}~\bibnamefont {Chen}}, \ and\ \bibinfo
  {author} {\bibfnamefont {C.~B.}\ \bibnamefont {Murray}},\ }\href
  {http://dx.doi.org/10.1038/nature08439} {\bibfield  {journal} {\bibinfo
  {journal} {Nature}\ }\textbf {\bibinfo {volume} {461}},\ \bibinfo {pages}
  {964} (\bibinfo {year} {2009})}\BibitemShut {NoStop}%
\bibitem [{\citenamefont {Callens}\ and\ \citenamefont
  {Zadpoor}(2018)}]{Callens2018}%
  \BibitemOpen
  \bibfield  {author} {\bibinfo {author} {\bibfnamefont {S.~J.~P.}\
  \bibnamefont {Callens}}\ and\ \bibinfo {author} {\bibfnamefont {A.~A.}\
  \bibnamefont {Zadpoor}},\ }\href
  {http://www.sciencedirect.com/science/article/pii/S1369702117306399}
  {\bibfield  {journal} {\bibinfo  {journal} {Materials Today}\ }\textbf
  {\bibinfo {volume} {21}},\ \bibinfo {pages} {241} (\bibinfo {year}
  {2018})}\BibitemShut {NoStop}%
\end{thebibliography}%
\end{document}